\documentclass[runningheads]{llncs}
\usepackage[T1]{fontenc}
\usepackage{graphicx}
\spnewtheorem{assumption}{Assumption}{\bfseries}{\itshape}

\spnewtheorem{conjecture2}{Conjecture}{\bfseries}{\itshape}

\usepackage{comment}
\usepackage{graphicx}
\usepackage{amsmath}
\usepackage{amsfonts}
\usepackage{amssymb}
\usepackage{mathrsfs}
\usepackage{braket}
\usepackage{bm}
\usepackage{multirow}
\usepackage{color}
\usepackage[normalem]{ulem}
\usepackage{mathrsfs}
\usepackage{mathtools}
\usepackage{float}
\usepackage{enumitem}
\usepackage{subfigure}
\usepackage{chngcntr}

\makeatletter

\newcommand{\Rmnum}[1]{\expandafter\@slowromancap\romannumeral #1@}
\makeatother

\def\OO{{\cal O}}

\def\ket#1{|\, #1\,\rangle}

\def\openone{{1\!\!1}}

\def\gg{\textsl{g}}

\def\NC1{{\bf NC}$^1$}

\begin{document}

\title{Encrypted Operator Computing: a novel scheme\\ for computation on
  encrypted data}
%
%
\author{Claudio Chamon\inst{1,2}
  \and
  Jonathan Jakes-Schauer\inst{2}
  \and
  Eduardo R. Mucciolo\inst{2}
  \thanks{Department of Physics, University of Central Florida
  (UCF). E. R. Mucciolo has conducted this research as part of a
  personal outside arrangement with USEncryption, Inc. The entity has
  licensed technology from UCF. The research and research results are
  not, in any way, associated with UCF.}
  \and
  Andrei E. Ruckenstein\inst{1,2}}
%
%
\institute{
  {Physics Department, Boston University, Boston,
    MA 02215, USA}
  \and
  {USEncryption Research, Orlando, FL 32826, USA}}
\maketitle              
\begin{abstract}
  We introduce a new approach to computation on encrypted data --
  Encrypted Operator Computing (EOC) -- as an alternative to Fully
  Homomorphic Encryption (FHE). Given a plaintext vector
  $\ket{x}$, $x\in \{0,1\}^n$, and a function $F(x)$ represented as an
  operator $\hat F$, $\hat F\;\ket{x} = \ket{F(x)}$, the EOC scheme is
  based on obfuscating the {\it conjugated operator} (circuit)
  $\hat{F}^E = \hat E\;\hat F\;\hat{E}^{-1}$ that implements
  computation on encrypted data, $\hat E \ket {x}$. 
  The construction of EOC hinges on the existence of a two-stage \NC1
  reversible-circuit-based IND-CCA2 cipher $\hat{E} = \hat{N} \hat{L}$, where
  $\hat{L}$ and $\hat{N}$ represent, respectively, linear and
  non-linear \NC1 tree-structured circuits of 3-bit reversible
  gates. We make and motivate security assumptions about such a \NC1
  cipher. Furthermore, we establish the polynomial complexity of the obfuscated
  circuit, the {\it evaluator} $O(\hat{F}^E)$, by proving that: (a)
  conjugation of each gate of $F$ with $\hat{L}$ yields a polynomial
  number of gates; and (b) the subsequent conjugation with $\hat{N}$
  yields a polynomial number of ``chips,'' $n$-input/$n$-output
  reversible functions, with outputs expressed as polynomial-sized
  ordered Binary Decision Diagrams (OBDDs). The security of individual
  chips is connected to the notion of Best Possible
  Obfuscators~\cite{Goldwasswer-Rothblum} which relies on poly-size
  OBDDs and the fact that OBDDs are normal forms that expose the
  functionality but hide the gate implementation of the chip. We
  conjecture that the addition of random pairs of NOTs between layers
  of $\hat{N}$ during the construction of $F^E$, a device analogous to
  the AddRoundKey rounds of AES, ensures the security of the evaluator.
  We also present a generalization to asymmetric
  encryption.

\end{abstract}
%
%
%



\section{Introduction}
\label{sec:introduction}

Current schemes for Fully Homomorphic Encryption (FHE) evolved from
Gentry's bootstrapping breakthrough~\cite{gentry2009}. Much progress
has been made improving the bootstrapping and the performance of
Somewhat Homomorphic Encryption (SWHE) schemes on top of which FHE is
constructed. These SWHE formulations involve lattice-based encryption
protocols~\cite{Regev-and-Co}, such as Learning With Errors
(LWE)~\cite{Regev}, ring-LWE~\cite{RLWE}, or
tori-LWE~\cite{Chillotti2020}. Despite the elegance of the approach
and the enormous effort directed towards building libraries that
implement both SWHE and FHE
schemes~\cite{MS-Seal,IBM-HELib,TFHE,Palisade,FV-NFLib,HEAAN}, there
are still severe limits to the practicality of these methods. For
example, addition and multiplication of 32- or 64-bit precision
numbers (messages) requires times of the order of
minutes~\cite{cryptoeprint_32_64_bit}. Moreover, the packing of
multiple messages into one ciphertext, which allows
parallelization~\cite{Smart2014}, presumes that data is pre-packed in
a specific way; if data needs to be collected from different
ciphertexts, the amortization gains from this type of vectorization
disappear. Given these practical limitations, it is valuable to
explore alternative approaches to FHE.


\section{Our Contribution}
\label{sec:contribution}

Here we present a different paradigm that we refer to as Encrypted
Operator Computing (EOC), in which operations on encrypted data are
carried out via an encrypted program - {\it the evaluator} - based on
reversible computation~\cite{Fredkin1982}. Reversible logic allows us
to formulate computation on encrypted data in terms of operators
(circuits of gates) in a transformed frame acting on transformed state
vectors (data). The change of frame hides information about both the
operators (the program) and the state (the data).

For the purposes of this paper, it is convenient to represent a
function $F$ as an operator $\hat F$ acting on a state vector
$\ket{x}$ associated to the binary data $x\in \{0,1\}^n$, such that
${\hat F}\,\ket{x} = \ket{F(x)}$.~\footnote{We concentrate on
  reversible functions $F$ as any function can be computed using
  reversible logic if one allows for the introduction of ancilla bit
  lines~\cite{nielsen2002quantum}, which are also included in the
  state vector.} Encryption is implemented by an operator $\hat E$
representing a permutation $E$ in $S_{2^n}$ that maps a plaintext
vector $\ket{x}$ onto a ciphertext vector $\ket{E(x)}$, with the
operator $\hat E^{-1}$ representing decryption.

An asymmetric encryption extension is enabled by a choice of
probabilistic encryption (which we use below) in combination with the
access to addition and multiplication operations on encrypted data,
with the exponentially many encryptions of 1 (unity) providing
multiple possible public keys.

In the operator language defined above, the encryption of $F(x)$ is
written as
\begin{align}
  \hat E\; \ket{F(x)}
  &=
    \hat E\;\hat F \;\ket{x}
    \nonumber\\
  &=
  \hat E\;\hat F \; \hat E^{-1}\;\hat E\;\ket{x}
  = \hat F^E \;\ket{E(x)}
    \;,
    \label{eq:EFx}
\end{align}
and thus the operator
\begin{align}
  \hat F^E \equiv \hat E\;\hat F\;\hat E^{-1}
  \;,
  \label{eq:F_E}
\end{align}
which represents what we refer to as {\it conjugation of the operator
$\hat F$ with the operator $\hat E$}, implements computation on
encrypted data. 

The EOC scheme involves two principal elements: the cipher, $\hat E$;
and the encrypted operator, $\hat F^E$. An adversary is presented with
encryptions of the data, $\ket{E(x)}$; and the evaluator,
$O(\hat F^E)$, an obfuscation of $\hat F^E$ that relies on the
specific structure of the cipher, $\hat E$ (see below). The
evaluator is represented as a concatenation of a polynomial number of
``chips'', $n$-input/$n$-output reversible functions, the outputs of
which are expressed as polynomial-sized ordered Binary Decision
Diagrams (OBDDs). OBDDs are normal forms that only expose the
functionality of the chip but hide its precise circuit
implementation. This paper provides proofs for the polynomial
complexity of the evaluator and connects the security of the scheme to
an extension of Best Possible Obfuscation introduced by Goldwasser and
Rothblum~\cite{Goldwasswer-Rothblum}. More precisely, we argue that
individual chips are realizations of Best Possible Obfuscators, and
that correlations among chips are erased by the random insertion of
NOT gates during the construction of the evaluator via the process of
conjugation. The random insertion of NOTs is analogous to the
AddRoundKey rounds of AES. The success of the EOC hinges on two
assumptions, which we motivate later in Sec.~\ref{sec:cipher} and
Sec.~\ref{sec:security}, namely: (i) that the cipher $\hat E$ is
secure to chosen plaintext/ciphertext attacks (more precisely, it
possesses the property of indistinguishability under adaptive chosen
ciphertext attacks -- IND-CCA2); and (ii) that the implementation of
the evaluator, $O(\hat F^E)$, obfuscates both $E$ and $F$.

\subsection{The cipher $E$:} 

\begin{figure}[h]
  \centering
  \vspace{0.5cm}
  \includegraphics[angle=0,scale=0.4]{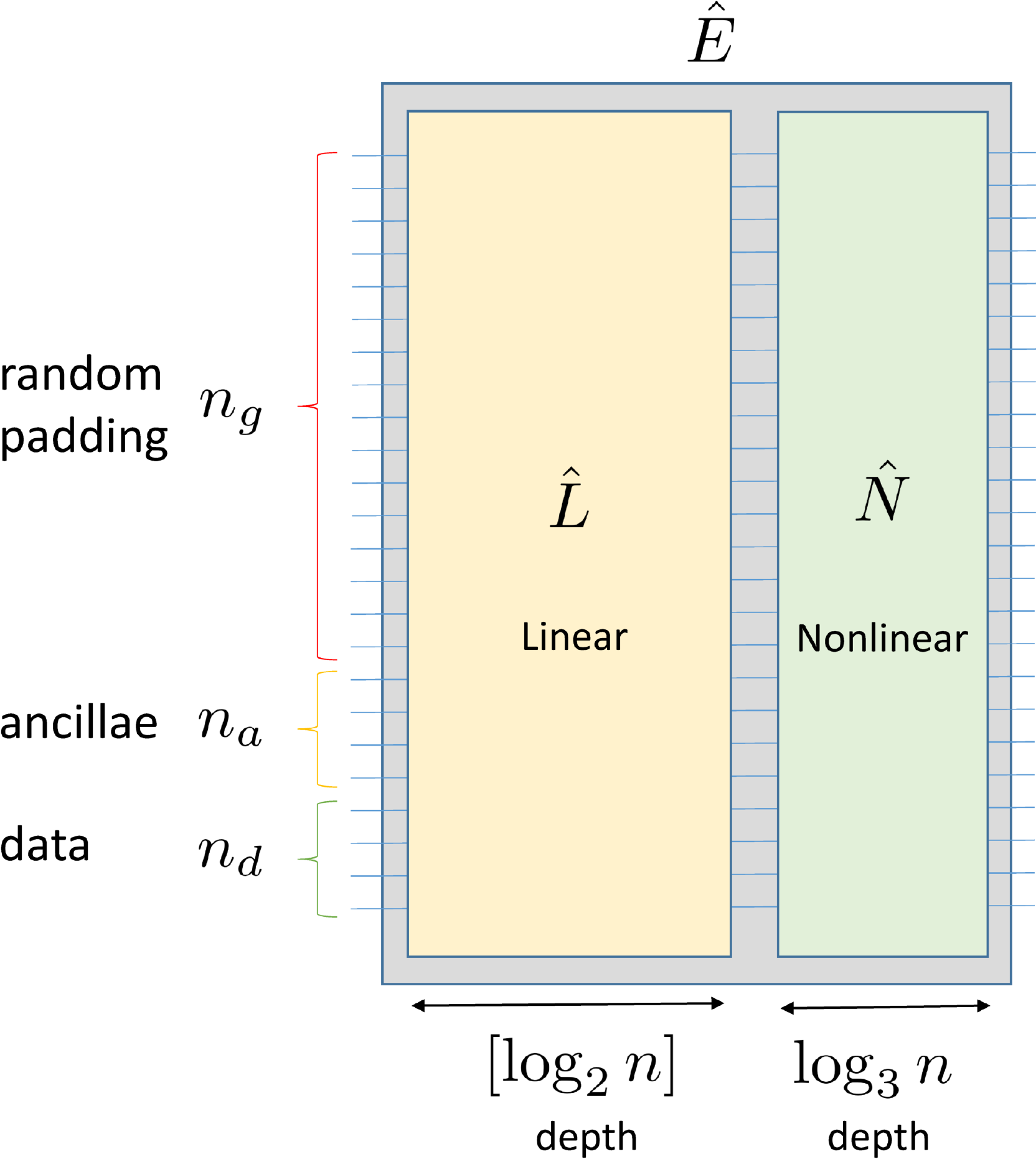}
  \vspace{0.5cm}
  \caption{Structure of the two-stage cipher consistent of $\log_2 n$
    layers of inflationary linear reversible 3-bit gates, followed by
    $\log_3 n$ layers of super-nonlinear reversible 3-bit gates. The
    cipher acts on $n$ bitlines, of which $n_d$, $n_a$, and $n_g$
    represent, respectively, data, ancillae, and random padding
    lines.}
  \label{fig:register_cipher}
\end{figure}

The scheme employs a 2-stage tree-structured (see
Sec.~\ref{sec:preliminaries}) reversible-circuit based cipher $E$ of
the form $\hat E=\hat N\;\hat L$, where {\it the linear stage}
$\hat L$ is implemented as a \NC1 circuit of gates drawn uniformly
from a set 144 linear inflationary 3-bit gates; and the {\it nonlinear
  stage} $\hat N$ is implemented as a \NC1 circuit of gates drawn
uniformly from 10752 super-nonlinear gates.  The cipher key $\kappa$
contains the list of which specific inflationary or super-nonlinear
gates are drawn. Multiple encryptions of the data, which is required
for enabling secure computation on encrypted data~\footnote{For
  example, exponentially many encryptions of Boolean {\tt False} and
  {\tt True} are needed to carry out secure Boolean computation on
  encrypted data.}, is realized by partitioning of the $n$-bit
register into $n_d$ bits of data ($x_d$), $n_a$ ancilla bits ($x_a$),
and the padding of the rest of the register by $n_g$ bits chosen
unifomly at random as 0 or 1 ($r$), as illustrated in
Fig.~\ref{fig:register_cipher}. (For concreteness, we fix the fraction
$n_g/n$ to be at least 2/3, so that each gate of the first layer of
$\hat L$ acts on a triplet containing two random bits of the padding
and one of data or ancillae.) Thus the $E(x)$ used throughout the
paper should be more precisely interpreted as
$E(x_d,x_a,r;\kappa)$. For each value of $x_d$ and $x_a$, there are
$2^{n_g}$ choices of $r$, hence exponentially many different
encryptions of a given datum.

The first assumption, critical for the polynomial implementation of
the EOC, is that the 2-stage \NC1 circuit $\hat E$ of the form
depicted in Fig.~\ref{fig:register_cipher} yields a secure
probabilistic encryption scheme. More precisely,
\begin{assumption}\label{A1}
  A 2-stage, tree-structured circuit, $\hat E=\hat N\;\hat L$, with
  one linear stage $\hat L$ comprised of $\log_2 n$ layers of linear
  gates drawn uniformly from the set of 144 linear inflationary gates,
  and a second nonlinear stage $\hat N$ comprised of $\log_3 n$ layers
  of nonlinear gates drawn uniformly from the set of 10752
  super-nonlinear gates, such that a fraction $2/3 < \eta <1$ of the
  $n$ input bits are drawn randomly from $\{0,1\}$ yields an IND-CCA2
  cipher.
\end{assumption}

The requirement of a shallow \NC1 cipher will be essential in ensuring
that the overhead of conjugation, which generically leads to an
exponential growth in complexity with the number of layers of
$\hat E$, remains polynomial in $n$. The special structure of the
2-stage cipher will be motivated by using the framework for analyzing
ciphers described in~\cite{cipher-paper}. There we introduced a
3-stage reversible-circuit based shallow cipher and used tools from
quantum statistical mechanics and connected plaintext and ciphertext
(higher-order) differential attacks to certain out-of-time-order
(OTOC) correlators and a residual (Pauli string) entropy, the
vanishing of which signal chaos and irreversibility in quantum
many-body systems. In Ref.~\cite{cipher-paper} we argued that the {\it
  exponential} vanishing of OTOCs and the saturation of the string
entropy are necessary conditions for the security of the block cipher.
 	
\subsection{The evaluator $O(\hat F^E)$:}
We are now in position to consider the essential elements of
conjugation of $\hat F$ with the cipher $\hat E = \hat N\,\hat L$ in
Assumption~\ref{A1}, which we implement in two steps:
$\hat F^L=\hat L\;\hat F\;\hat L^{-1}$, followed by 
$\hat F^E = \hat N\;\hat F^L\;\hat N^{-1}$. As already alluded to above, a third step in building the evaluator, which ensures the security of the scheme, is the random insertion of random NOTs between layers of $\hat N$ in the course of the second step of conjugation (with $\hat N$).   

\subsubsection*{(i) Conjugation of $\hat F$ with the linear
  stage $\hat L$}:

\vspace{0.2cm} The implementation of EOC proceeds by (i) decomposing
$\hat F$ as a circuit of $M$ elementary gates (NOTs, CNOTs, and
Toffoli gates), $\hat F = \hat f_M\cdots \hat f_2\,\hat f_1$; and (ii)
carrying out the conjugation operation with $\hat L$ gate-by-gate (of
$\hat F$):
\begin{align}
  \hat F^L
  &=
    \hat L\;\hat f_M\cdots \hat f_2\,\hat f_1\;\hat L^{-1}
  \nonumber\\
  &=
    \left(\hat L\; \hat f_M\;\hat L^{-1}\right)\,
    \cdots
    \left(\hat L\; \hat f_2\;\hat L^{-1}\right)\,
    \left(\hat L\; \hat f_1\;\hat L^{-1}\right)
    \nonumber\\
  &=
    \hat f^L_M\cdots \hat f^L_2\,\hat f^L_1
  \;,
  \label{eq:breaking_F_L}
\end{align}
where each
$\hat f^L_i\equiv \hat L\; \hat f_i\;\hat L^{-1} =
\hat\gg_{i,Q_i}\dots \hat\gg_{i,2}\;\hat\gg_{i,1}$ is a circuit of
$Q_i$ elementary gates $\hat\gg_{i,q}$, $q=1,\dots, Q_i$. The
conjugation of $\hat f_i$ with the linear layers of gates of $\hat L$
is derived by applying ``collision rules'' for reversible
gates. Collision rules (presented in
Appendix~\ref{sec:conjugation_inflationary_gates}) reflect the fact
that, generally, elementary gates do not commute and that
interchanging the order of two gates generates additional ``debris''
gates.  In Sec.~\ref{sec:linear} below we argue that conjugation
with $\hat L$ leads to a proliferation of the number of
gates, $Q_i$, and, most importantly, scatters these gates across the
$n$ bitlines, thus diluting information about the initial gate
location, $\hat f_i$. We also prove:
\begin{theorem}
  Conjugation of a gate $\hat f_i$ (a NOT, CNOT, or Toffoli gate) with
  $\hat L$ yields a worse-case overhead factor of
  ${Q_i^{\rm max}}\le n^{\mu_3}$, and average overhead factor of
  $\overline{Q}_i\le n^{\nu_3}$ (averaged over circuits $\hat L$), for
  every elementary gate $\hat f_i$ of $\hat F$, with
  ${\mu_3}=3\;\log_2 3 \approx 4.75$ and
  ${\nu_3}=3\;\log_2 \frac{7}{3}\approx 3.67$.
  \label{thm:1}
\end{theorem}
This theorem places a rigorous polynomial upper bound on both the
maximum and the average number of gates (where the average is over
linear circuits $\hat L$) that result from this conjugation process.

\subsubsection*{(ii) Conjugation of $\hat F^L$ with the nonlinear
  stage $\hat N$}:

\vspace{0.2cm}

The next step in the construction is the conjugation of each of the
$n_C=\sum_{i=1}^M Q_i$ gates $\hat\gg_{i,m}$ in the product
$\hat f^L_M\cdots \hat f^L_2\,\hat f^L_1$ with the $\OO(\log n)$
layers of the nonlinear operator $\hat N$:
\begin{align}
  \hat F^E
  &=
    \hat N\;\hat f^L_M\cdots \hat f^L_2\,\hat f^L_1\;\hat N^{-1}
    \nonumber\\
  &=
    \hat N\;\hat\gg_{M,Q_M}\dots \;\hat\gg_{1,2}\;\hat\gg_{1,1}
    \;\hat N^{-1}
    \nonumber\\
  &=
    \left(\hat N\, \hat\gg_{M,Q_M}\,\hat N^{-1}\right)\,
    \cdots
    \left(\hat N\, \hat\gg_{1,2}\,\hat N^{-1}\right)\,
    \left(\hat N\, \hat\gg_{1,1}\,\hat N^{-1}\right)
    \nonumber\\
  &=
    \hat\gg^N_{M,Q_M}\dots \;\hat\gg^N_{1,2}\;\hat\gg^N_{1,1}
  \;.
  \label{eq:breaking_F_N}
\end{align}
The end result is the collection of $n_C$ ``chips'',
$\hat\gg^N_{i,q} \equiv \hat N\;\hat\gg_{i,q}\;\hat N^{-1}$, the gate
make-up of which is obfuscated by expressing the $n$ outputs of every
chip as Ordered Binary Decision
Diagrams~\cite{Bryant_review,Knuth-book} (OBDDs or simply BDDs for
short). The evaluator $O(\hat F^E)$ is the concatenation of these
BDD-expressed chips. In Sec.~\ref{sec:nonlinear}, we prove a
polynomial bound for the number of nodes of the BDDs, thus establishing that the chip
BDDs are polynomial-sized OBDDs (POBDDs):
\begin{theorem}
  A chip seeded by conjugation of a gate $\hat\gg_{i,q}$ (a NOT, CNOT,
  or Toffoli gate) with $\hat N$ yields BDDs with at most $n^{\gamma}$
  nodes for each of the $n$ outputs of the chip, where
  $\gamma=\log_3 7\approx 1.77$.
  \label{thm:2}
\end{theorem}

Since a BDD is a normal form representing all Boolean
functions of the same functionality, the resulting $n$ POBDDs provide
a concise representation of the chip $\hat\gg^N_{i,q}$ that exposes no
more information than necessary to recover the chip's
functionality. For individual chips, this last step realizes the Best
Possible Obfuscation via POBDDs introduced by Goldwasser and
Rothblum~\cite{Goldwasswer-Rothblum}.

We re-emphasize that the two-stage process outlined above only
requires polynomial overhead, as it yields a polynomial number of
POBDDs. This hinges on the $\OO(\log n)$ depth of each of the two
stages: the shallow depth ensures that the linear stage leads to a
polynomial number of gates, and that the nonlinear stage produces
POBDDs for each of those gates. The complexity of the EOC is
determined by the polynomial expansion factor due to the conjugation
with $\hat L$ and the polynomial sizes of the BDDs resulting from
conjugation with $\hat N$.

The overall time complexity of EOC (per gate of $\hat F$) is bounded
by $n^{\mu_3+2}$ or $n^{\mu_3+1}$ if the BDDs of a chip are evaluated
in series or in parallel, respectively. The space required, as
measured by the number of nodes in all BDDs, is bounded by
$n^{\mu_3+\gamma+1}$.

\subsubsection*{(iii) Injection of randomness during
  conjugation with $N$}:

\vspace{.2cm}

Since the form of each chip, namely,
$\hat\gg^N_{i,q} = \hat N\;\hat\gg_{i,q}\;\hat N^{-1}$, is known to an
adversary, one should ask whether Best Possible is ``Good Enough'' in
ensuring that one cannot walk back from the output BDDs and learn
information essential for decryption. In Sec.~\ref{sec:security} we
argue that the structure of conjugation leads to complete erasure of
information on $\hat N$ in a ``dark zone'', in which gates in $\hat N$
that do not overlap with the central gate $\hat\gg_{i,q}$ annihilate
their mirror-inverse in $\hat N^{-1}$. Even in the region beyond the
``dark zone'' -- the ``light cone'' seeded by $\hat\gg_{i,q}$ --
information is partially erased due to the specific structure of
conjugation, i.e., there are exponentially many possibilities for
$\hat\gg_{i,q}$ and the gates in $\hat N$ and $\hat N^{-1}$ that yield
the same chip. In this sense, Best Possible Obfuscation is, indeed,
good enough in obfuscating individual chips.

However, the result of conjugation of the full function $\hat F^E$
involves the concatenation of a large (but polynomial) number of
chips. Best Possible Obfuscation cannot be applied to the full
function $\hat F^E$, since combining multiple chips into one would
lead to exponential-size BDDs for the $n$ output lines of the full
computation. Is then Best Possible Obfuscation of individual chips
sufficient to guarantee the obfuscation of a concatenation of multiple
chips? Can an adversary with access to the whole collection of chips
extract and integrate information from correlations among BDDs
representing multiple chips? Can information erased in one chip
become visible in another?~\footnote{One may ask whether
  correlations among the $n$ output BDDs for a single chip may reveal
  more information about the cipher than a BDD for a single
  output. Since for a single chip the information lost through the
  ``dark zone'' is the same for all outputs (because it is seeded by
  the same initial gate), this question is not as relevant as that for
  the case of multiple chips.}
  
In the absence of a concrete way of addressing these questions, we
take advantage of the freedom of inserting random identities in the
form of random pairs of NOTs which are then distributed across the
system between conjugation with consecutive layers of $\hat
N$. Because of the injection of random pairs of NOTs between chips,
construction of chips must be carried in parallel, for each layer of
$\hat N$ (as discussed Sec.~\ref{sec:security}). This randomization
process leaves the sizes of chip BDDs unchanged, but scrambles the
functionality of individual chips while preserving the functionality
of the concatenation of chips representing the entire function. The
addition of randomness washes out correlations among chips and confers
to the full evaluator security beyond that provided by the Best
Possible Obfuscation of individual chips. In fact, here we make a
stronger assumption, namely:

\begin{assumption}\label{A2}
  Best Possible Obfuscation of each chip of a collection of chips
  injected with random pairs of NOTs hides both $\hat N$ and the seed
  gates, $\hat\gg_{i,q}$, thus yielding an Indistinguishability
  Obfuscation of the conjugated circuit $\hat F^E$.
\end{assumption}

In Sec.~\ref{sec:security} we motivate this conjecture heuristically
by discussing how BDDs are altered by the randomness, and by
drawing an analogy between the insertions of the random NOTs between
layers of $\hat N$ and the AddRoundKey rounds of AES.

In summary, we propose EOC as a novel reversible-logic-based approach
to computation on encrypted data. In this paper we prove the
polynomial overhead of the EOC scheme, the security of which is tied
to Assumptions 1 and 2.

\subsection{Organization of the paper}

This paper is organized as follows. We start in
Sec.~\ref{sec:preliminaries} with a more detailed description of the
tree-structured circuits and the 3-bit gate sets required to implement
the cipher $E$. In Sec.~\ref{sec:cipher} we motivate
Assumption~\ref{A1} by using the framework of the quantum statistical
mechanics approach to encryption introduced in
Ref.~\cite{cipher-paper}. In Sec.~\ref{sec:evaluator} we present the
details of the implementation of conjugation with the linear
($\hat L$) and nonlinear ($\hat N$) stages of the cipher, and prove
Theorems~\ref{thm:1} and \ref{thm:2}. The security of the scheme and
the motivation of Assumption~\ref{A2} are discussed in
Sec.~\ref{sec:security}.  Finally, conclusions and a discussion of
future directions are presented in Sec.~\ref{sec:conclusions}. A brief
introduction to reversible computing, and to conjugation rules that
follow from the non-commutatitivity of gate operators are given in
Appendix~\ref{sec:tools}; other relevant details are presented in
Appendices~\ref{sec:conjugation_inflationary_gates}
and~\ref{linear_network_model_bound}.


\section{Preliminaries: Tree-structured Circuits}
\label{sec:preliminaries}

Below we describe in detail circuits mentioned in the introduction, in
which triplets of bits acted upon by 3-bit gates are arranged in a
hierarchical (tree) structure. We consider the case when $n$ is a
power of 3, $n=3^q$. We proceed by forming groups of triplets of
indices for each layer, selected as follows:
\begin{align}
  \ell = 1: & \quad(0,1,2)\;(3,4,5)\;(6,7,8) \dots \nonumber\\
  \ell = 2: & \quad(0,3,6)\;(1,4,7)\;(2,5,8) \dots \nonumber\\
  \ell = 3: & \quad(0,9,18)\;(1,10,19)\;(2,11,20) \dots \nonumber\\
  \ell = 4: & \quad(0,27,54)\;(1,28,55)\;(2,29,56) \dots \nonumber\\
  \dots \quad&
\end{align}
More precisely, each of the $n/3=3^{q-1}$ triplets in layer $\ell$ are
indexed by $(i,j,k)$, which we write in base 3 as
\begin{align}
  i=&z_0 + 3\;z_1+3^2\;z_2+\dots+3^{\ell-1}\times\underline{0}+\dots\;3^{q-1}\;z_{q-1}\nonumber\\
  j=&z_0 + 3\;z_1+3^2\;z_2+\dots+3^{\ell-1}\times\underline{1}+\dots\;3^{q-1}\;z_{q-1}\nonumber\\
  k=&z_0 + 3\;z_1+3^2\;z_2+\dots+3^{\ell-1}\times\underline{2}+\dots\;3^{q-1}\;z_{q-1}
      \;,
      \label{eq:trit-triples}
\end{align}
where $z_a=0,1,2$, for $a=0,\dots, q-1$. Notice that at layer $\ell$
the members of the triplets, $(i,j,k)$, are numbers that only differ
in the $(\ell-1)$-th trit, while the other $q-1$ trits
$z_a, a\ne \ell-1$, enumerate the $3^{q-1}=n/3$ triplets. (If more
than $q$ layers are needed, we recycle in layer $\ell > q$ the
triplets of layer $\ell\!\!\mod q$.)

Once the triplets of indices, $(i,j,k)$, are selected for each layer,
we map them onto groups of three bits indexed by,
$\left(\pi(i),\pi(j),\pi(k)\right)$, via a (randomly chosen)
permutation $\pi$ of the $n$ bitlines. We note that for these
tree-structure ciphers, the key consists of the data needed to specify
the circuit, namely: (i) the permutation $\pi$; and (ii) the list of
gates in $S_8$ chosen to act on each of the triplets, for all
layers. This key uniquely defines the circuit, and its inverse.

\section{Assumption~\ref{A1}: motivation of the
  \NC1 cipher}
\label{sec:cipher}

Posing Assumption~\ref{A1} is motivated by
the work in Ref.~\cite{cipher-paper}, where we proposed the 3-stage
reversible circuit in \NC1 as a specific
realization of pseudo-random permutations. There we used ideas and
tools from quantum statistical mechanics to connect the security of
block ciphers to arbitrary-order differential attacks to quantitative
measures used to diagnose irreversibility and chaos in the dynamics of
quantum circuits.

Most importantly, in Ref.~\cite{cipher-paper} we argued that the
specific 3-stage design is necessary in order to eliminate polynomial
tails in a certain {\it stay probability for short Pauli strings},
which would translate into vulnerability to polynomial differential
attacks. Eliminating these tails with generic (unstructured)
circuit-based ciphers would require more than $\OO(\log n)$ layers, a
conclusion that is consistent with both the work on the minimum depth
required for scrambling quantum circuits~\cite{Harrow2018approximate}
and the discussion of pseudo-random functions in {{\bf NC}$^{i+1}$,
  $i\ge 1$} presented in~\cite{naor-reingold}.

The intuition that explains the special properties of the 3-stage
cipher of Ref.~\cite{cipher-paper} is based on two elements: (i)
the segregation of linear and non-linear gates into separate stages;
and (ii) the use of special gate sets in constructing both the linear
and non-linear stages of the cipher. The linear ``inflationary'' gates
in $\hat L_{l,r}$ flip two bits of the output for a single bit-flip in
the input, thus accelerating the spreading of the effect of a
single-bit-flip across the $n$ bitlines of the circuit. The
super-nonlinear gates in $\hat N$ maximize production of
(Pauli)-string entropy, which restricts the information an adversary
can extract from plaintext/ciphertext attacks. Both linear and
nonlinear stages must be $\OO(\log n)$-depth circuits in order to
entwine all bitlines. The role of the interplay between linear and
nonlinear stages in determining the exponential security of the
3-stage cipher is illustrated by the quantitative discussion of the
Strict Avalanche Criterion (SAC) in Ref.~\cite{cipher-paper}. In
particular, the special choice of inflationary gates leads to a
double-exponential decay of the SAC with the number of applied linear
layers, but only after an initial decay due to the production of
string entropy induced by the nonlinear layers. (For
$\ell\sim \log n$, a double exponential in $\ell$ translates into an
exponential in $n$.)

Assumption~\ref{A1} adds probabilistic encryption to the scheme and
removes one of the linear stages of the discussion in ref.~\cite{cipher-paper}. Probabilistic encryption yields multiple
encryptions of a given datum, a property needed in order to implement
secure computation on encrypted data. Fig.~\ref{fig:register_cipher}
illustrates the 2-stage architecture of the cipher
$\hat E = \hat N\;\hat L$ with the partition of the $n$-bit register
into $n_d$ bits of data, $n_a$ ancilla bits, and the padding of the
rest of the register by $n_g$ bits chosen randomly as 0 or 1. (For
concreteness, we fix the fraction $n_g/n$ to be at least 2/3, so that
each gate of the first layer of $\hat L$ acts on a triplet containing
two random bits of the padding and one of data or ancillae. This
connectivity accelerates the randomizing action of the cipher, and
could be used to optimize EOC.) To motivate why padding the plaintext
with random bits allows us to remove the linear stage on the
ciphertext side, let us consider the SAC for both plaintext and
ciphertext attacks, using the analytical tools developed in
Ref.~\cite{cipher-paper}.

Starting from the ciphertext side, one first passes (from right to
left) through the nonlinear stage, which is responsible for extensive
entropy production and an initial exponential decay of the SAC
correlator with the number of layers in $\hat N$, $\ell_N$. Seeded by
the initial decay, the action of the linear stage (again traversed
from right to left) leads to a double exponential decay of the SAC
with the number of layers of $\hat L$, $\ell_L$. Consequently, for
$\ell_L=\log_2 n$ the SAC correlator describing ciphertext attacks
decays exponentially in $n$.

From the plaintext side, the injection of randomness into an extensive
fraction of input bits leads to a depression of the SAC correlator on
those bits, which then propagates across all bitlines and becomes
doubly-exponentially small in $\ell_L$ through the action of the
layers of $\hat L$. We note that the SAC can already be suppressed
exponentially (in $n$) past a single layer of inflationary gates
designed such that each gate in that layer acts on one bit of data (or
ancilla) and on two bits of the random padding. The propagation
proceeds through the nonlinear layers (from left to right), the action
of which leads to the saturation of the string entropy while building
the essential nonlinearity of the encryption process.

We remark that even though we used the SAC to guide the above
discussion, a picture based on string inflation and proliferation
described in Ref.~\cite{cipher-paper} can be used to argue more
generally that the 2-stage cipher in Fig.~\ref{fig:register_cipher} is
as secure to plaintext arbitrary differential attacks as the 3-stage
cipher of Ref.~\cite{cipher-paper}. We cannot argue the same for
ciphertext attacks, as the string entropy becomes extensive but not
necessarily maximal.

The arguments above lead us to Assumption~\ref{A1}, namely that the
2-stage cipher with added random padding possesses the property of
indistinguishability under adaptive chosen ciphertext attack
(IND-CCA2). We note that this is a weaker assumption than the one we
would advance for the 3-stage cipher, which we conjecture implements a
strong pseudo-random permutation.

\section{Construction of the evaluator, $\hat F^E$}
\label{sec:evaluator}

Here we detail the two steps involved in building the evaluator: the conjugation with the linear stage of the cipher, $\hat L$: $\hat F^L=\hat L\;\hat F\;\hat L^{-1}$, and the proof of Theorem~\ref{thm:1} are presented in Sec.~\ref{sec:linear}; and the subsequent conjugation with the nonlinear stage, $\hat N$,  
$\hat F^E = \hat N\;\hat F^L\;\hat N^{-1}$, and the proof of Theorem~\ref{thm:2} are presented in Sec.~\ref{sec:nonlinear}. 

\subsection{Conjugation by linear gates}
\label{sec:linear}

The conjugations with $\ell$ layers of linear inflationary gates are
implemented gate-by-gate. Conjugation by a single inflationary gate
involves interchanging gates according to commutation rules derived
from circuit equivalences (which we refer to as ``collisions'') and
then applying simplifications that follow from operator identities
derived via simple Boolean algebra. Appendix~\ref{sec:tools} presents
a self-contained discussion of collisions and simplifications that
allow us to derive the rules for conjugation of elementary gates (NOT,
CNOT, or Toffoli) with each linear inflationary gate of each layer of
$\hat L$.

\begin{figure}[t]
  \centering
  \vspace{0.5cm}
  \includegraphics[width=\textwidth,angle=0,scale=0.95]{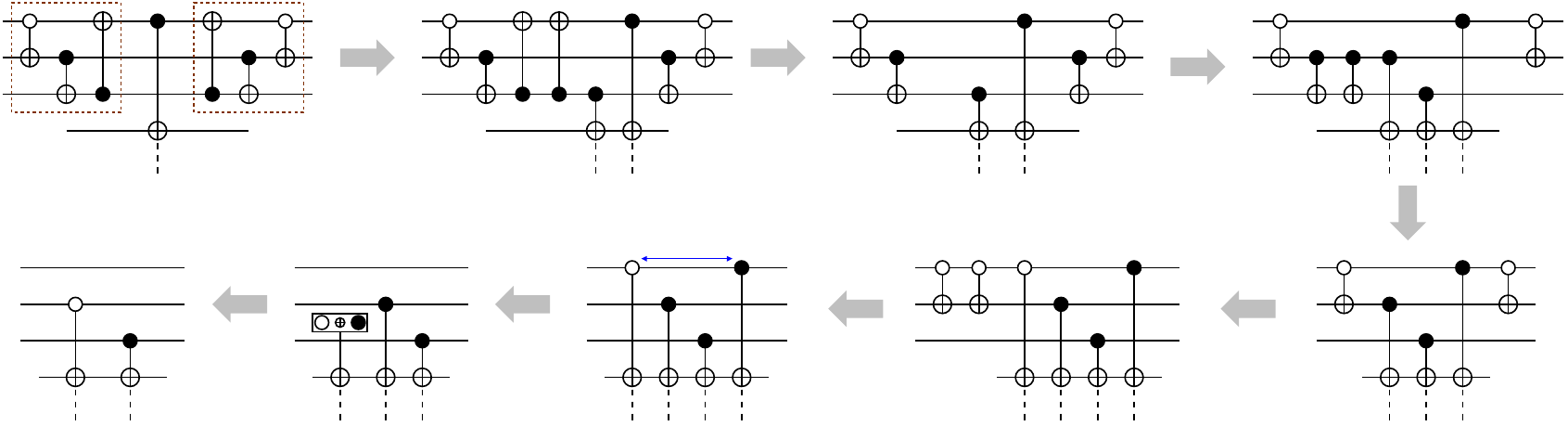}
  \vspace{0.5cm}
  \caption{An example of conjugation of a controlled gate by an
    inflationary gate depicted as the dashed box containing CNOTs (see
    Appendix~\ref{sec:conjugation_inflationary_gates} for the
    decomposition of the 144 inflationary gates in terms of CNOTS). In
    the case shown, a control bit of the gate being conjugated
    overlaps with one of the 3 bitlines of the inflationary gate. The
    equivalent circuit is obtained through a sequence of substitutions
    according to the collision and simplification rules in
    Appendices~\ref{sec:collision_rules} and~\ref{sec:simplification}.
    Notice that, as a result of conjugation, memory of the control bit
    (overlapping with the inflationary gate) of the original
    (``mother'' gate) is lost and two bitlines acquire controls
    associated with two offspring gates.}
  \label{fig:inflationary_1_bit_conjugation_control}
\end{figure}

As an illustration,
Fig.~\ref{fig:inflationary_1_bit_conjugation_control} shows the
conjugation of a gate with a control bit overlapping with one of the
three bitlines of the inflationary gate and its inverse, which are
both broken up into CNOTs of both polarities. The dashed line
indicates that other possible control and the target bitline of the
gate to be conjugated lie outside of the three bitlines of the
inflationary gate and its inverse. By alternating between collision
and simplification rules in Appendix~\ref{sec:collision_rules}
and~\ref{sec:simplification}, one arrives at the two gates in
Fig.~\ref{fig:inflationary_1_bit_conjugation_control}; the controls of
both of these gates touch bitlines different from that touched by the
control of the original gate. These two gates commute since their
other controls and target (those attached to the dashed lines) act
identically on the same bitlines.

Following the procedure illustrated above, we derive the set of rules
describing conjugation by all types of inflationary gates and all
possible overlap configurations of their bitlines with the target and
controls of the gate being conjugated. These rules are summarized in
Figs.~\ref{fig:conj_general1}, \ref{fig:conj_general2},
\ref{fig:conj_general3}, and \ref{fig:conj_general4} of
Appendix~\ref{sec:conjugation_inflationary_gates}. (The example worked
out explicitly in
Fig.~\ref{fig:inflationary_1_bit_conjugation_control} above
corresponds to case C4 of Fig.~\ref{fig:conj_general3}.) These rules
are not exhaustive: for nearly every case there is a plurality of
other equivalent but different configurations of offspring gates. For
instance, one can flip polarities of controls using the polarity
mutation rules of Appendix~\ref{sec:simplification}.

  
The conjugation through multiple layers of inflationary gates, which
follows from the recursive application of the conjugation rules from
Figs. \ref{fig:conj_general1}, \ref{fig:conj_general2},
\ref{fig:conj_general3}, and \ref{fig:conj_general4}, should be viewed
as a branching process, with controls and targets scattering and
touching an increasing number of bitlines as more layers of
inflationary gates are deployed. This process increases the number of
gates, but each of these gates has no more controls than the original
gate, a consequence of the linearity of inflationary gates. The growth
in the number of gates and the scattering of targets and controls
across all bitlines of the circuits leads to ambiguity about the
specific gate $\hat f_i$ that is being conjugated. The example of
Fig.~\ref{fig:inflationary_1_bit_conjugation_control} already
illustrates the mechanism for this behavior: one original gate splits
into two offsprings, with their respective controls scattering
elsewhere.

We note that since NOTs and CNOTs are linear gates, their conjugation
with $\hat L$ yields a linear circuit, which could be easily
synthesized directly~\footnote{Linearity allows the synthesis of a
  reversible circuit using the outputs resulting from only $n+1$
  inputs, e.g., $x=0$ and $x=1,2,\dots,2^{n-1}$. For every input in
  this $(n+1)$-long list, one builds the correct output (without
  changing outputs from previous inputs in the list) by using $\OO(n)$
  linear gates. Therefore one can synthesize any linear reversible
  circuit with at most $\OO(n^2)$ NOTs and CNOTs.}. For an initial NOT
gate, the resulting circuit would contain only NOTs, touching on
average $n/2$ bitlines for $\hat L$ sufficiently deep
($\ell = \log_2 n$ suffices, as shown in
Ref.~\cite{cipher-paper}). Similarly, conjugation of an initial CNOT
would yield a generic linear circuit, which can be synthesized with
at most $\OO(n^2)$ (CNOT and NOT) gates.

These arguments cannot be applied to the conjugation of the nonlinear
Toffoli gate, in which case we can derive a bound on the proliferation
of gates through the explicit use of the conjugation rules in
Appendix~\ref{sec:conjugation_inflationary_gates}.

\subsection*{Proof of Theorem~\ref{thm:1}}
\label{sec:linear_bound}

\begin{figure}[h]
  \centering
  \vspace{0.5cm}
  \includegraphics[angle=0,scale=0.55]{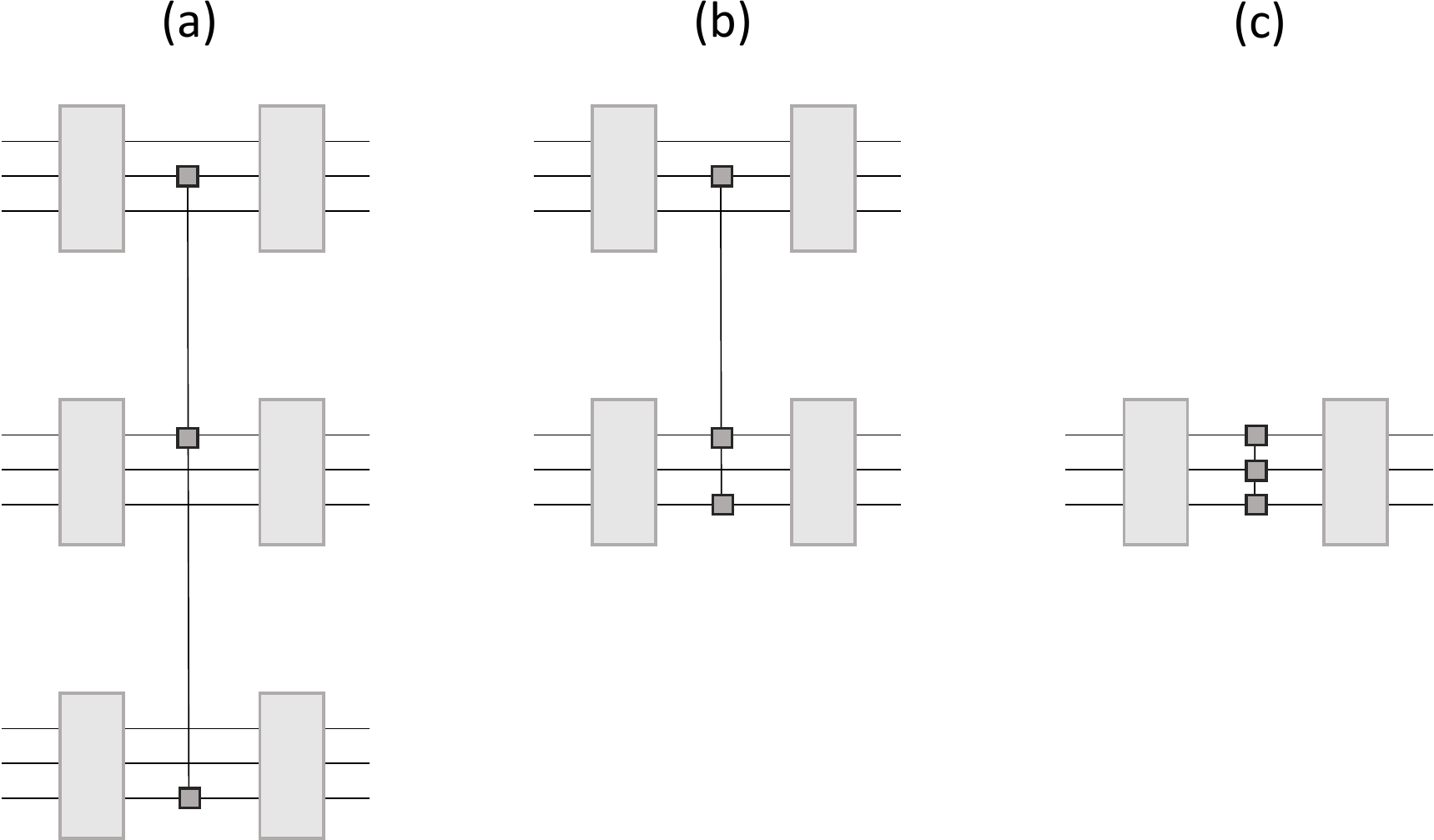}
  \vspace{0.5cm}
  \caption{Three arrangements of overlaps between the three bitlines
    touched by a Toffoli gate and inflationary gates (on the left
    side) and their inverses (on the right side), represented as gray
    boxes. The bitlines of the Toffoli are represented by dark gray
    squares that, for the purpose of illustrating the three
    arrangements, do not distinguish between controls and target. The
    three arrangements correspond to cases where: (a) each bitline of
    the Toffoli gate overlaps with a different inflationary gate in a
    layer of $\hat L$; (b) two of the bitlines of the Toffoli gate
    overlap with one inflationary gate, while the remaining bitline
    overlaps with another inflationary gate; and (c) all three
    bitlines of the Toffoli gate overlap with the same inflationary
    gate.}
  \label{fig:arrangement}
\end{figure}

We separate three arrangements, illustrated in
Fig.~\ref{fig:arrangement}, corresponding to whether one, two, or
three inflationary gates in a layer of $\hat L$ overlap with the three
bitlines covered by the Toffoli gate.

\begin{figure}[h]
  \centering
  \vspace{0.5cm}
  \includegraphics[angle=0,scale=0.55]{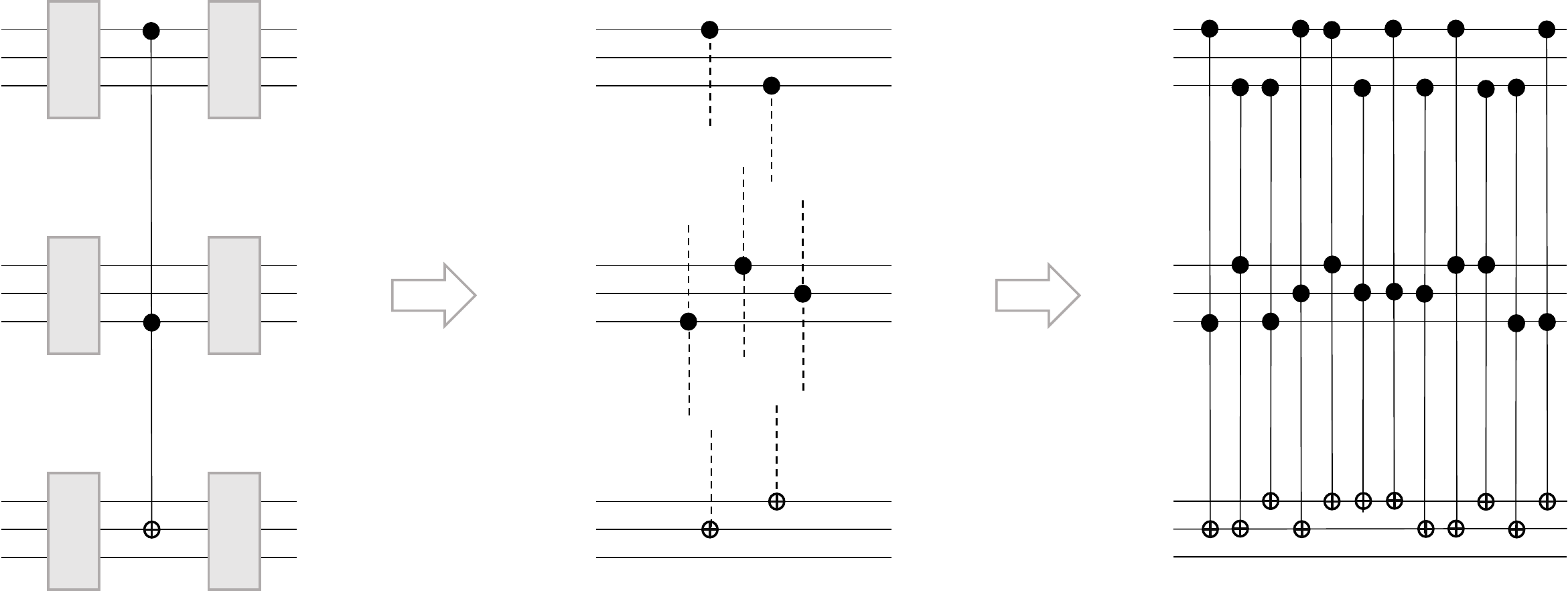}
  \vspace{0.5cm}
  \caption{An example of conjugation of a Toffoli gate overlapping with
    three different inflationary gates. The process generates
    $2\times 3\times 2$ Toffoli gates as follows: (i) conjugation of a
    control by one inflationary gate (on top) results in 2 possible
    control positions; (ii) conjugation of a control with a second
    inflationary gate (in the middle) results in 3 possible control
    positions; and finally, (iii) conjugation of a target with a third
    inflationary gate (at the bottom) results in 2 possible target
    positions. The 12 resulting Toffoli gates encompass all ways of
    choosing two controls and one target, each one from different
    groups of bitlines associated with each of the three different
    inflationary gates.}
    \label{fig:example-3inflationary}
\end{figure}

In the arrangement where three inflationary gates overlap with the
Toffoli gate, each will cover separately one of the three bitlines of
the gate, as shown in Fig.~\ref{fig:arrangement}a. The conjugations in
cases A1-A6, B1-B6, C1-C6, and D1-D6 in
Figs.~\ref{fig:conj_general1},~\ref{fig:conj_general2},~\ref{fig:conj_general3},
and \ref{fig:conj_general4}, respectively, describe the independent
scattering of the controls and target of the Toffoli gate. In all
these instances, a control or target overlapping with one inflationary
gate scatters into either 2 (in 2/3 of cases) or 3 (in 1/3 of cases)
controls or targets as a result of conjugation. New Toffoli gates are
generated by this process, corresponding to all possible choices of
groupings of two controls and one target, each of which is picked from
the set of possibilities generated by separate conjugation with each
of the three different inflationary gates. (The counting is made
easier by considering the conjugation by the three inflationary gates
one at a time.) An example is illustrated in
Fig.~\ref{fig:example-3inflationary}. In this arrangement the maximum
number of Toffoli gates that can be generated is $3^3$. The average
number is $[2\times 2/3 + 3\times 1/3]^3= (7/3)^3$.

The next arrangement we consider (see Fig.~\ref{fig:arrangement}b) is
one in which two inflationary gates overlap with the three bitlines of
the Toffoli gate: one inflationary gate overlaps with two of those
three bitlines, and the other with only one. For the inflationary gate
that overlaps with only one of the three bitlines, whether the bitline
contains a control or target, the scattering possibilities are the
same as those considered above. The possibilities resulting from the
conjugation with the other inflationary gate, which overlaps with two
bitlines of the Toffoli gate, are summarized in cases A7-A15, B7-B15,
C7-C15, and D7-D15 of
Figs.~\ref{fig:conj_general1},~\ref{fig:conj_general2},~\ref{fig:conj_general3},
and \ref{fig:conj_general4}, respectively. Notice that all these
scatterings produce at most 4 possibilities, which is less than or
equal to the number of possibilities that would be generated by
conjugation with two {\it independent} inflationary gates as in the
previous arrangement, namely $2\times 2, 2\times 3, 3\times 2$ or
$3\times 3$. Therefore, both the maximum and the average number of
Toffoli gates that can be generated are less than or equal to the
values obtained in the arrangement in Fig.~\ref{fig:arrangement}a,
namely $3^3$ for the maximum and $(7/3)^3$ for the average.

Finally, for the arrangement in Fig.~\ref{fig:arrangement}c, in which
all the bitlines of the Toffoli gate falls within those covered by a
single inflationary gate, we need to consider the cases A16-A18,
B16-B18, C16-C18, and D16-D18 of
Figs.~\ref{fig:conj_general1},~\ref{fig:conj_general2},~\ref{fig:conj_general3},
and \ref{fig:conj_general4}, respectively. The largest number of
Toffoli gates generated by conjugation is 7, which is smaller than the
minimum number of gates $2\times 2\times 2$ that would be generated in
the arrangement of Fig.~\ref{fig:arrangement}a. In this case, the
maximum and the average number of Toffoli gates that can be generated
are (again) less than or equal to $3^3$ for the maximum and $(7/3)^3$
for the average.

We therefore conclude that, per layer of conjugation with $\hat L$,
the number of Toffoli gates is increased by a factor of no more than
$3^3$, and on average no more than $(7/3)^3$. Conjugation with $\ell$
such layers yield expansion factors
\begin{align}
  {Q}^{\rm max}\le 
    3^{3\ell}
    \quad
    \text{and}
    \quad
    \overline{Q}\le 
    \left(\frac{7}{3}\right)^{3\ell}
    \;.
    \label{eq:N-Toffoli-bound}
\end{align}
For $\ell = \log_2 n$, we find that ${Q}^{\rm max}\le n^{\mu_3}$ and
$\overline{Q}\le n^{\nu_3}$, where the exponents are
${\mu_3}=3\;\log_2 3 \approx 4.75$ and
${\nu_3}=3\;\log_2 \frac{7}{3}\approx 3.67$, thus proving
Theorem~\ref{thm:1}.

\subsection{Conjugation by nonlinear gates}
\label{sec:nonlinear}

Each of the elementary gates (NOTs, CNOTs, and Toffoli gates) of the
circuit resulting from conjugation with the linear stage $\hat L$ of
the cipher is then conjugated with the remaining, nonlinear stage
$\hat N$ of the cipher, according to Eq.~\eqref{eq:breaking_F_N}. We
cast the reversible circuit resulting from the conjugation by
nonlinear gates as a collection of chips,
$\hat\gg^N_{i,q} \equiv \hat N\;\hat\gg_{i,q}\;\hat N^{-1}$, defined
in detail below.

A chip implements a reversible computation represented by the
reversible function $h(x)$, where $x$ is an $n$-bit input and $h(x)$
is the $n$-bit output. The binary function $h_i(x)$ corresponds to the
$i$-th output bit of $h(x)$, and can be encoded as a BDD. The function
$h_i(x)$ may not depend on all the $n$ input bits, but instead its
domain is a subset $b[h_i]$ of those inputs; we denote as the width of
the $h_i$ BDD the cardinality, $|b[h_i]|$, of that set. The footprint
$b[h]$ of the chip $h$ is the union
$b[h]\equiv b[h_0]\;\cup\; b[h_1]\;\cup\;\cdots\;\cup \;b[h_{n-1}]$,
and the width of the chip the cardinality $|b[h]|$.

Let us consider the conjugation of the chip $h$ by a 3-bit nonlinear
gate $g$, starting with the BDD representation of the Boolean
functions $h_i(x), i=0,\dots, n-1$. Our aim is to obtain the BDD
representation of the Boolean functions $h^g_i(x), i=0,\dots, n-1$,
where $h^g(x)\equiv g\left(h(g^{-1}(x)\right)$ is the result of the
conjugation of $h$ by $g$. The gates $g,g^{-1}\in S_8$ act on three
bits labeled by $j_1<j_2<j_3$, and their action can be expressed as
three Boolean output functions, $g_{j_1}, g_{j_2}, g_{j_3}$ and
$g^{-1}_{j_1}, g^{-1}_{j_2}, g^{-1}_{j_3}$. The Boolean expression for
$h^g_i(x)$ is constructed in two steps:
\begin{align}
  &\text{step 1}:\label{eq:step1}
  \\
  &\quad \tilde h_{i}(x) \equiv
  h_i\left(x_0,\dots,
  x_{j_1}\leftarrow g^{-1}_{j_1}(x_{j_1}, x_{j_2}, x_{j_3}),\dots,
  x_{j_2}\leftarrow g^{-1}_{j_2}(x_{j_1}, x_{j_2}, x_{j_3}),\dots,
  x_{j_3}\leftarrow g^{-1}_{j_3}(x_{j_1}, x_{j_2}, x_{j_3}),\dots,
    x_{n-1}\right)
    \nonumber\\
  \nonumber\\
  \nonumber\\
  &\text{step 2}:\label{eq:step2}
  \\
  &\quad h^g_{i}(x) \equiv
    \begin{cases}
      \tilde h_i(x),& i\notin \{j_1,j_2,j_3\}\\
      g_i\left(\tilde h_{j_1}(x), \tilde h_{j_2}(x), \tilde h_{j_3}(x)\right),
      & i\in \{j_1,j_2,j_3\}\;.\nonumber\\
      \end{cases}
\end{align}
Starting with BDDs expressing the $h_i, i=0,\dots,n-1$, one constructs
the BDDs for the $\tilde h_i, i=0,\dots,n-1$ of step 1 by using the
{\it composition} rules for BDD manipulation~\cite{Bryant1986}, and
from those one proceeds to construct the BDDs for the
$h^g_{i}, i=0,\dots,n-1$ using the {\it apply} rules~\cite{Bryant1986}.

These operations of gate conjugation through BDD manipulation are
carried out for all gates in a layer of nonlinear gates. The procedure
is then iterated for all layers of the nonlinear circuit $\hat N$. At
the end of the process, we have a reversible operator encoded as a
vector of (at most $n$) BDDs -- the chip. The footprint of the chip
grows with the number of layers of conjugation, and so do the sizes of
the BDDs, i.e, the number of terminal and non-terminal nodes of the
BDDs contained in the chip. The size of the chip is defined as the
size of the largest BDD in the chip. We remark that the sizes of BDDs
depend on the input variable order, and choosing different orders for
the BDDs associated to each output reduces the chip size. In proving
Theorem~\ref{thm:2} we use different variable orders for different
outputs. (We will comment on the bound for the case with the same
variable order at the end of the section.)


\subsection*{Proof of Theorem~\ref{thm:2}}
\label{sec:scalingBDD}

We proceed with the proof of Theorem~\ref{thm:2} by first proving a
Lemma for the conjugation of a NOT with the $\log_3 n$ layers of
nonlinear gates of the tree-structured circuit $\hat N$.

\begin{lemma}
  A chip seeded by conjugation of a NOT gate with $\hat N$ yields at
  most $n^{\gamma}$ nodes, with $\gamma=\log_3 7\approx 1.77$, for
  each BDD representing one of the $n$ outputs of the chip.
\label{lemma:NOT}
\end{lemma}

\begin{proof}
  
Suppose the NOT gate acts on bitline $t$. When this NOT gate is
sandwiched between 3-bit gates $g,g^{-1}$, one obtains a 3-bit
permutation that acts on a triplet of bits
$\left(\pi(i_0),\pi(i_1),\pi(i_2)\right)$, where $i_{z_0}$ is obtained
from $i=\pi^{-1}(t)$ according to the tree structure, described in
Ref.~\cite{cipher-paper}, by replacing its least significant trit by
$z_0=0,1,2$ (notice that one of $i_0,i_1,i_2$ must be equal to
$i$). Each of the three output bits is a Boolean function represented
by a BDD of footprint $x_{\pi(i_0)},x_{\pi(i_1)}$ and $x_{\pi(i_2)}$,
of width 3. Upon conjugating with the second layer, the width of the
chip increases to 9, encompassing the bits
$\pi(i_{z_0z_1}), z_0,z_1=0,1,2$, with the index $i_{z_0z_1}$ obtained
by substituting the two least significant trits of $i$ by $z_0$ and
$z_1$. Continuing along this path, after the $\ell$-th layer the chip
will have grown to width $3^\ell$, encompassing bits
$\pi(i_{z_0z_1\dots z_{\ell-1}}), z_0,z_1,\dots,z_{\ell-1}=0,1,2$,
where $i_{z_0z_1\dots z_{\ell-1}}$ are obtained by manipulating the
first $\ell$ trits of $i$. We note that the tree-like growth of the
chip described above, and illustrated in Fig.~\ref{fig:tree}, ensures
that every bitline covered at level $\ell$ of the conjugation scheme
is always accompanied by two freshly touched bitlines at the next
level, $\ell+1$.

  \begin{figure}[h]
  \centering
  \vspace{0.5cm}
  \includegraphics[angle=0,scale=0.85]{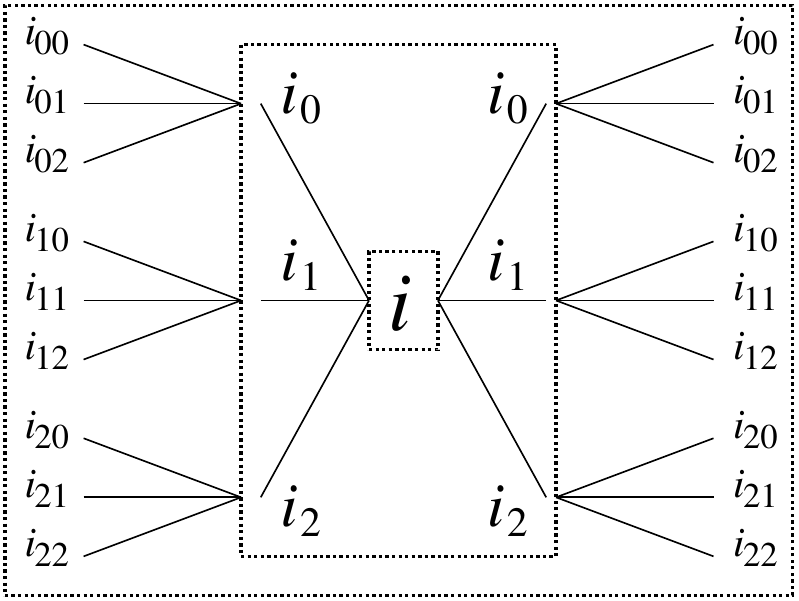}
  \vspace{0.5cm}
  \caption{Hierarchical chip construction. In the example, the chip is
    seeded by a 1-bit NOT gate at bitline $t=\pi(i)$. (For simplicity,
    we take the permutation $\pi$ to be the identity in this example.)
    At level $\ell=0$ the chip has a 1-bit footprint containing only
    $i$. At level $\ell=1$, the footprint encompasses 3 bitlines,
    $i_0,i_1$ and $i_2$, which are obtained from $i$ by replacing its
    lowest significant trit by 0,1, and 2, respectively. Notice that
    one of $i_{z_0}, z_0=0,1,2$, equals $i$ itself; the other two are
    fresh bitlines, accreted to the footprint. At level $\ell=2$, the
    chip encompasses 9 bitlines, $i_{z_0z_1}, z_0,z_1=0,1,2$, which
    are obtained from the 3 bitlines of the previous level by
    replacing the second lowest significant trit $z_1$ in each of
    $i_0, i_1$ and $i_2$ by $z_1=0,1,2$. One of the
    $i_{z_0z_1}, z_1=0,1,2$ equals $i_{z_0}$, while the other two
    values of $z_1$ correspond to the fresh bitlines added to chip,
    for each of $z_0=0,1,2$. The recursion proceeds similarly for
    levels $\ell>2$.}
  \label{fig:tree}
\end{figure}

The scaling of the size of the BDDs associated with the $3^\ell$
outputs of the chip after $\ell$ layers of conjugation can also be
obtained recursively. As illustrated in Fig.~\ref{fig:BDDs}a, we start
with the BDD for the NOT gate, which has one non-terminal node with
the variable $x_{\pi(i)}$ and the two terminal nodes, $\top$ and
$\bot$, using Knuth's notation of Ref.~\cite{Knuth-book} for true and
false, respectively. The BDDs resulting from conjugation with the
first layer, which touches the bitline $\pi(i)$ via a single gate,
$g$, results in a chip with three outputs,
$h^g_{\pi(i_0)},h^g_{\pi(i_1)}$ and $h^g_{\pi(i_2)}$, each encoded in
a BDD with three inputs, $x_{\pi(i_0)},x_{\pi(i_1)}$ and
$x_{\pi(i_2)}$. These BDDs can be constructed following the
prescription above. The first step is the calculation of
$\tilde h_{\pi(i)}$ through the substitution
$x_{\pi(i)}\leftarrow g_{\pi(i)}^{-1}(x_{\pi(i_{0})}, x_{\pi(i_{1})},
x_{\pi(i_{2})})$ [see Eq.~\eqref{eq:step1}]. This corresponds to the
replacement of the single, non-terminal node $\pi(i)$ in
Fig.~\ref{fig:BDDs}a by the non-terminal nodes of a BDD involving
three variables: the original $x_{\pi(i)}$ and the two fresh variables
that appear in the triplet with bitline ${\pi(i)}$ (recall that one of
$i_0,i_1$ or $i_2$ equals $i$). In Fig.~\ref{fig:BDDs}b we illustrate
this substitution with the worst-case scenario in which the function
$g_{\pi(i)}^{-1}$ is represented by a BDD with 7 non-terminal nodes,
the maximum size BDD on three variables. We remark that the other two
$\tilde h$ functions, expressing the outputs of the two fresh bitlines
involved in the triplet with ${\pi(i)}$ (two of
${\pi(i_{0})}, {\pi(i_{1})}$, and ${\pi(i_{2})}$) simply equal the
corresponding output bits from
$g^{-1}(x_{\pi(i_{0})}, x_{\pi(i_{1})}, x_{\pi(i_{2})})$, as they are
not affected by the original NOT gate.

The next step is to implement the calculation of
$h^g_{\pi(i_{z_0})}(x)= g_{\pi(i_{z_0})}\left(\tilde
  h_{\pi(i_0)},\tilde h_{\pi(i_1)},\tilde h_{\pi(i_2)}\right)$,
$z_0=0,1,2$, as prescribed by Eq.~\eqref{eq:step2}. Notice that the
$\tilde h$ functions associated with the two fresh bitlines (other
than $\pi(i)$) are already expressible using
$x_{\pi(i_{0})}, x_{\pi(i_{1})}, x_{\pi(i_{2})}$, and hence in
transforming from the BBD for $\tilde h_{\pi(i)}$ to the BDDs for
$h^g_{\pi(i_{z_0})}$, $z_0=0,1,2$ requires no additional non-terminal
nodes beyond the maximum 7 for a 3-variable BDD. While this statement
that the BDDs for $\tilde h$ and $h^g$ have comparable sizes is
trivial for conjugation with the first layer, it has important
implications for conjugation with subsequent layers. To retain this
property, we must order the input variables to the BDDs. In
particular, to prepare the 3-bit chip for conjugation with the second
layer, the input variable $x_{\pi(i_{z_0})}$ must appear last in the
BDD for the output $h^g_{\pi(i_{z_0})}$, $z_0=0,1,2$. More
generically, at any level of conjugation, the BDD expressing the
output of the chip on a given bitline must have the input variable on
that same bitline appear last, in preparation for the subsequent level
of conjugation.

  \begin{figure}[h]
  \centering
  \vspace{0.5cm}
  \includegraphics[angle=0,scale=0.25]{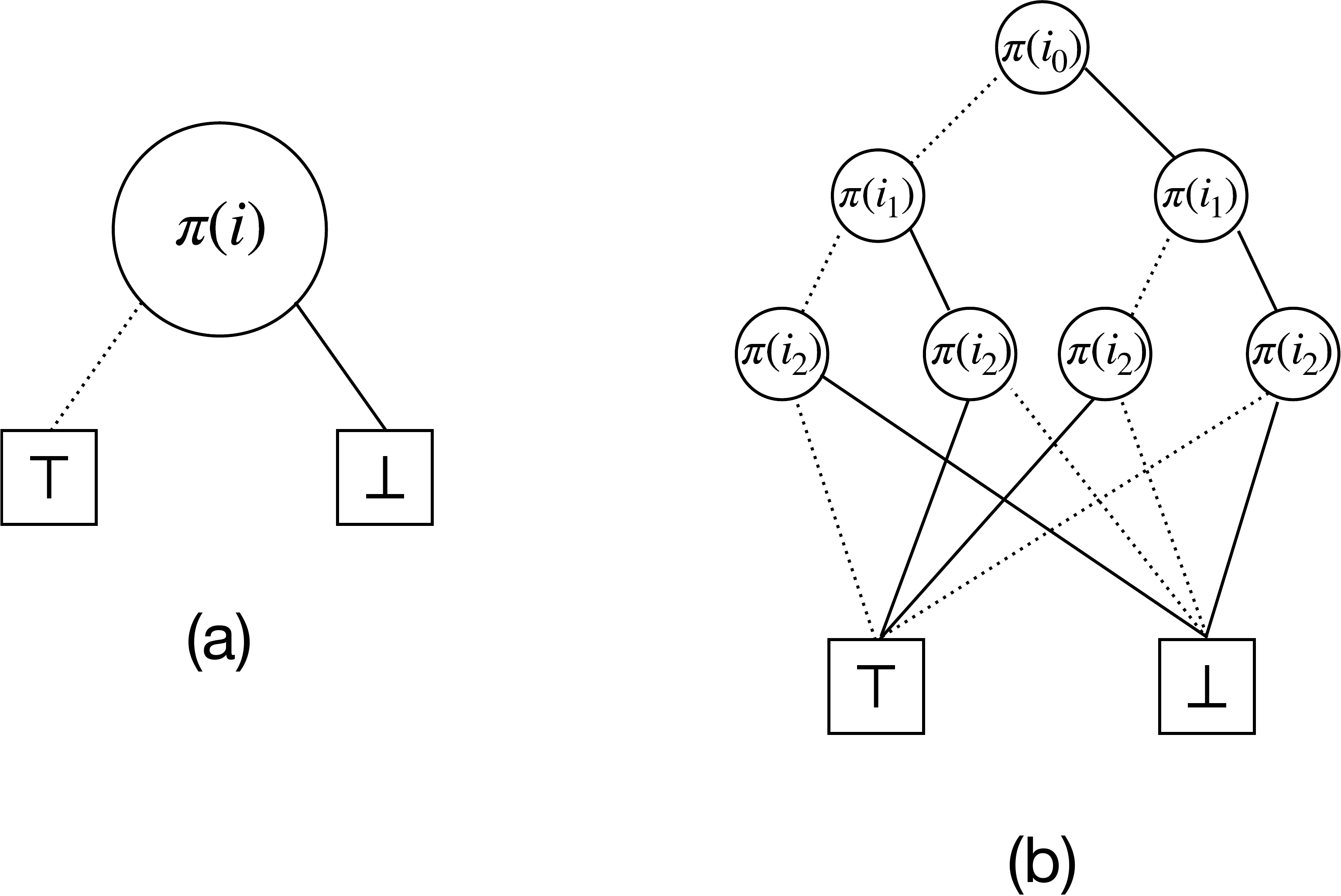}
  \vspace{0.5cm}
  \caption{(a) BDD for a NOT gate acting on bitline $t=\pi(i)$,
    corresponding to the $\ell=0$ seed for the chip in
    Fig.~\ref{fig:tree}. (b) At the next level, $\ell=1$, the
    substitution
    $x_{\pi(i)}\leftarrow g_{\pi(i)}^{-1}(x_{\pi(i_{0})},
    x_{\pi(i_{1})}, x_{\pi(i_{2})})$ replaces the node labeled by
    $\pi(i)$ in (a) by a BDD on variables $x_{\pi(i_0)}$,
    $x_{\pi(i_1)}$, and $x_{\pi(i_2)}$, corresponding to the triplet
    of bitlines ${\pi(i_0)}$, ${\pi(i_1)}$, and ${\pi(i_2)}$ on which
    a 3-bit gate in the first layer of the nonlinear operator $\hat N$
    overlaps with the chip at level $\ell=0$, at $\pi(i)$ (see
    Fig.~\ref{fig:tree}). We show a worst-case scenario, in which case
    7 non-terminal nodes are needed to represent the function
    $g_{\pi(i)}^{-1}$.}
  \label{fig:BDDs}
\end{figure}

We proceed with conjugation by the three gates in the second layer,
$g', g''$ and $g'''$, which overlap separately with bits ${\pi(i_0)}$,
${\pi(i_1)}$ and ${\pi(i_2)}$, respectively (see Fig.~\ref{fig:tree}). As
already described above, the tree structure of the cipher implies that
each of these three gates adds two fresh variables accompanying each
of the bitlines activated by the first layer.  Following the first
step described in Eq.~\eqref{eq:step1}, each of 
$\tilde h_{\pi(i_{z_0})}, z_0=0,1,2$, is implemented via the three
substitutions,
\begin{align}
  x_{\pi(i_0)}&
  \leftarrow {g'}_{\pi(i_0)}^{-1}(x_{\pi(i_{00})}, x_{\pi(i_{01})}, x_{\pi(i_{02})})
  \nonumber\\
  x_{\pi(i_1)}&
  \leftarrow {g''}_{\pi(i_1)}^{-1}(x_{\pi(i_{10})}, x_{\pi(i_{11})}, x_{\pi(i_{12})})
  \label{eq:subs}              
\\
  x_{\pi(i_2)}&
  \leftarrow {g'''}_{\pi(i_2)}^{-1}(x_{\pi(i_{20})}, x_{\pi(i_{21})}, x_{\pi(i_{22})})
  \;.
  \nonumber
  \end{align}
  (Again, notice that one of the indices $i_{z_00}, i_{z_01}$ or
  $i_{z_02}$ is the same as the original $i_{z_0}$.) The substitution
  amounts to replacing the non-terminal nodes $\pi(i_{0})$,
  $\pi(i_{1})$, and $\pi(i_{2})$ by small BDDs for the functions
  ${g'}_{\pi(i_0)}^{-1}$, ${g''}_{\pi(i_1)}^{-1}$, and
  ${g'''}_{\pi(i_2)}^{-1}$, respectively. Each replacement of
  $\pi(i_{z_0})$ by BDDs with nodes $\pi(i_{z_00}), \pi(i_{z_01})$ and
  $\pi(i_{z_02})$, $z_0=0,1,2$, leads to an increase in the total
  number of nodes of the BDDs for the
  $\tilde h_{\pi(i_{z_0})}, z_0=0,1,2$. In the worst-case scenario,
  these substitutions inflate the number of nodes by a factor of 7
  (the maximum number of non-terminal nodes in a BDD on three
  variables). It is critical to note that, as a consequence of the
  tree structure, this inflation happens independently for each of the
  three nodes $\pi(i_{z_0})$, $z_0=0,1,2$, and thus, the overall
  increase of the BDDs for $\tilde h_{\pi(i_{z_0})}, z_0=0,1,2$, is
  additive instead of multiplicative.

The three 9-variable $\tilde h_{\pi(i_{z_0})}, z_0=0,1,2$, were
constructed from the $h^g_{\pi(i_{z_0})}$ of the previous level of
conjugation, where the variable $x_{\pi(i_{z_0})}$ appears last in the
corresponding BDD. Thus, through the substitutions in
Eq.~\eqref{eq:subs}, $x_{\pi(i_{z_00})}, x_{\pi(i_{z_01})}$, and
$x_{\pi(i_{z_02})}$ are the last 3 variables appearing in the BDDs for
$\tilde h_{\pi(i_{z_0})}, z_0=0,1,2$. Moreover, these three
$\tilde h_{\pi(i_{z_0})}$ are each accompanied by two $\tilde h$
functions that only depend on the same 3 variables, and represent the
outputs associated with the two fresh bitlines involved in the triplet
with ${\pi(i_{z_0})}$. The second step which completes the conjugation
with the second layer, as prescribed in Eq.~\eqref{eq:step2}, is to
build
\begin{align}
  h^{g'}_{\pi(i_{0z_1})}(x)&= g'_{\pi(i_{0z_1})}\left(\tilde
  h_{\pi(i_{00})},\tilde h_{\pi(i_{01})},\tilde
  h_{\pi(i_{02})}\right)
  \nonumber\\
  h^{g''}_{\pi(i_{1z_1})}(x)&= g''_{\pi(i_{1z_1})}\left(\tilde
  h_{\pi(i_{10})},\tilde h_{\pi(i_{11})},\tilde
  h_{\pi(i_{12})}\right)
  \nonumber\\
  h^{g'''}_{\pi(i_{2z_1})}(x)&= g'''_{\pi(i_{2z_1})}\left(\tilde
  h_{\pi(i_{20})},\tilde h_{\pi(i_{21})},\tilde
  h_{\pi(i_{22})}\right)
  \;,
\end{align}
where $z_1=0,1,2$. Each of the arguments for each of the three
equations above contain one of the 9-variable
$\tilde h_{\pi(i_{z_0})}$ along with its two companion 3-variable
$\tilde h$s. Because the 3 variables in the two 3-variable $\tilde h$s
always appear last in the BDD for the 9-variable $\tilde h$, no new
nodes are required to build the BDDs for $h^{g'},h^{g''}$ and
$h^{g''}$, in the worst-case scenario in which the substitutions
involve 7 non-terminal nodes, again, the maximum for a 3-variable BDD.

Finally, in preparation for conjugation with the next layer, we order
the last 3 variables of each of the $h^{g'},h^{g''}$ and $h^{g''}$ so
that $x_{\pi(i_{0z_1})}$, $x_{\pi(i_{1z_1})}$, and $x_{\pi(i_{2z_1})}$
appear, respectively, as the last variable of the BDDs describing
$h^{g'}_{\pi(i_{0z_1})}$, $h^{g''}_{\pi(i_{1z_1})}$, and
$h^{g'''}_{\pi(i_{2z_1})}$, $z_1=0,1,2$. In other words,
$x_{\pi(i_{z_0z_1})}$, $z_0,z_1=0,1,2$, are placed as the last
variables of the BDDs for their corresponding output bitlines,
${\pi(i_{z_0z_1})}$, of the 9-bit chip.

These steps can be repeated for conjugation with the subsequent
layers: as illustrated in Fig.~\ref{fig:BDD_recursion}, nodes
$\pi(i_{z_0z_1\dots z_{\ell-1}})$ are substituted by a BDD with nodes
labeled as $\pi(i_{z_0z_1\dots z_{\ell-1}z_\ell})$, $z_\ell=0,1,2$
that represent a function $g^{-1}_{\pi(i_{z_0z_1\dots z_{\ell-1}})}$
of three variables $x_{\pi(i_{z_0z_1\dots z_{\ell-1}z_\ell})}$,
$z_\ell=0,1,2$. The figure displays the worst-case scenario, in which
7 non-terminal nodes replace the original node. Note that the LO and
HI branches~\cite{Knuth-book} of the substituted node
$\pi(i_{z_0z_1\dots z_{\ell-1}})$ are replaced, respectively, by the
branching lines terminating at the $\bot$ and $\top$ nodes of the
substituted BDD. As with previous layers, the general iteration
proceeds with the second step of conjugation, Eq.~\eqref{eq:step2},
followed by the reordering that places the input variable on each
bitline as the last one in that line's output BDD.

  \begin{figure}[h]
  \centering
  \vspace{0.5cm}
  \includegraphics[angle=0,scale=0.12]{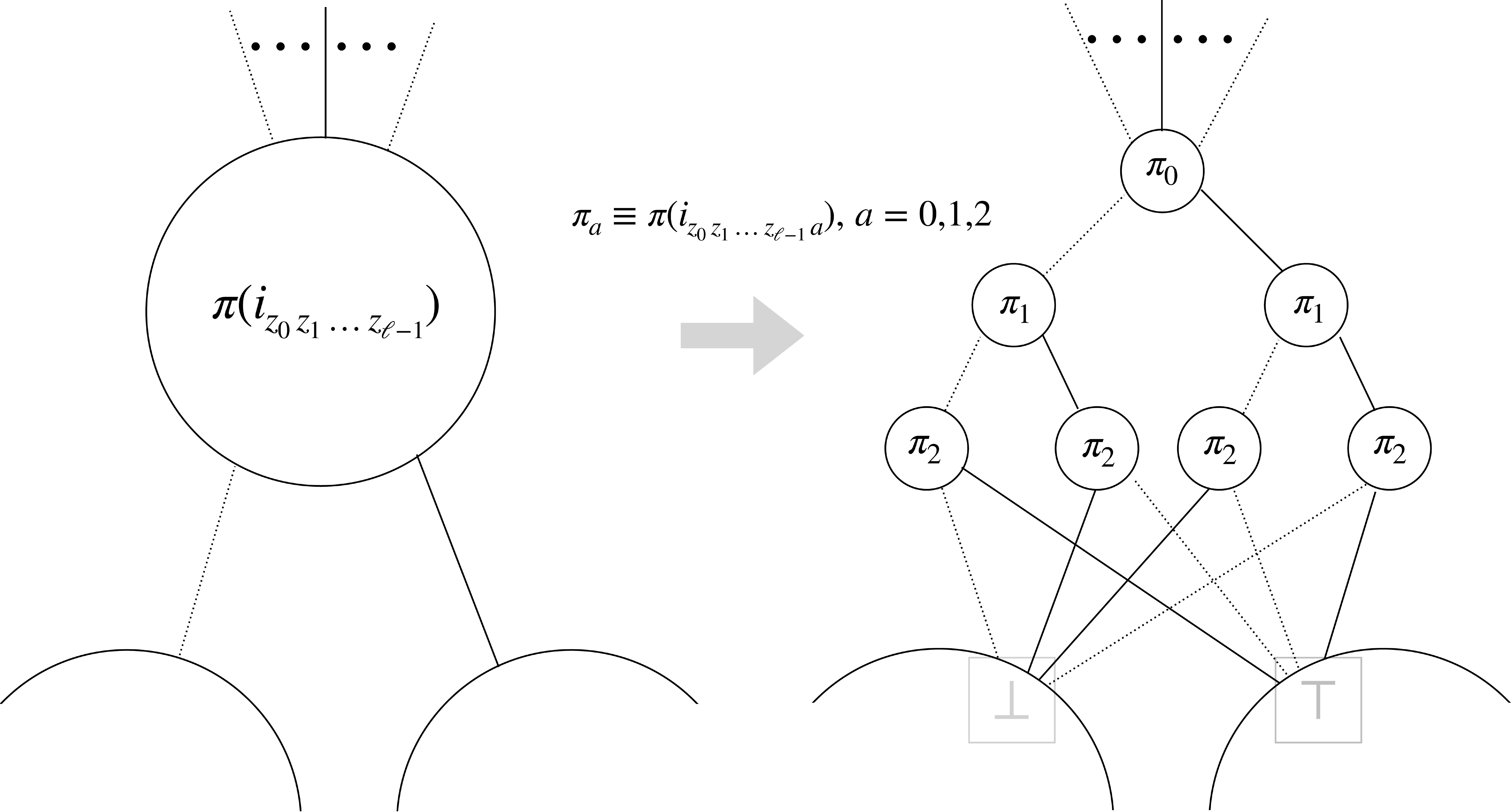}
  \vspace{0.5cm}
  \caption{Generalization of the substitution of a node by a BDD in
    the course of conjugation by a layer of $\hat N$. The node indexed
    by the bitline $\pi(i_{z_0z_1\dots z_{\ell-1}})$ is substituted by
    a BDD with nodes labeled as
    $\pi(i_{z_0z_1\dots z_{\ell-1}z_\ell})$, $z_\ell=0,1,2$, that
    represent a function $g^{-1}_{\pi(i_{z_0z_1\dots z_{\ell-1}})}$ of
    three variables $x_{\pi(i_{z_0z_1\dots z_{\ell-1}z_\ell})}$,
    $z_\ell=0,1,2$. Again, the example shows a worst-case scenario
    with 7 non-terminal nodes needed to represent the function
    $g^{-1}_{\pi(i_{z_0z_1\dots z_{\ell-1}})}$. The LO and HI branches
    of the substituted node (left) are replaced, respectively, by the
    branching lines terminating at the $\bot$ and $\top$ nodes (shown
    in light gray) of the substituted BDD (right).}
  \label{fig:BDD_recursion}
\end{figure}

The above construction, based on the tree structure of the cipher,
allows us to place bounds on the size of the BDDs describing the
outputs of the chip, after conjugation with $\ell$ layers. If the BDDs
for each output of the chip have size bounded by $B_{\rm max}(\ell)$
at level $\ell$, then at level $\ell+1$ the maximum size satisfies the
recursion
\begin{align}
  B_{\rm max}(\ell+1)-2
  &=
  7\;\left[B_{\rm max}(\ell)-2\right]
  \;,
  \label{eq:max_size_recursion}
\end{align}
which reflects the increase in the number of non-terminal nodes by a
factor of at most 7. (There are always two terminal nodes, hence the
subtraction of 2 on both sides.) Seeding the recursion with
$B_{\rm max}(0)=3$ (the size of the BDD representing a simple NOT
operation) yields
\begin{align}
  B_{\rm max}(\ell)
  =
  7^\ell + 2
  \;.
  \label{eq:max_size_ell}
\end{align}
In Appendix~\ref{linear_network_model_bound} we present an alternative
derivation of this bound that uses the linear network model of
computation explained by MacMillan~\cite{MacMillan},
Bryant~\cite{Bryant_review}, and Knuth~\cite{Knuth-book}.



After conjugation with all the $\ell=\log_3 n$ layers of our cipher,
the footprint of the BDD saturates, covering all of the $n$
bitlines. For this value of $\ell$ one reaches
\begin{align}
  B_{\rm max}(\log_3 n)
  =
  n^{\log_3 7} + 2
  \sim
  n^{1.7712}
  \;.
  \label{eq:max_size_n}
\end{align}
Adding the sizes of the BDDs for all the output bits of the chip
yields its volume $V(\ell)$, i.e., the total number of terminal and
non-terminal nodes needed to represent all the outputs of the
chip. For the same value of $\ell$ that saturates the footprint, i.e.,
$3^\ell=n$, one obtains the bound on the volume,
\begin{align}
  V_{\rm max}(\log_3 n)
  =
  n\;\left(n^{\log_3 7} + 2\right)
  \sim
  n^{2.7712}
  \;.
  \label{eq:max_volume_n}
\end{align}

This completes the proof of Lemma \ref{lemma:NOT}.

\vspace{.5cm}

We remark that the bound on the size of the BDDs remains polynomial in
the case where we adopt the same order of variables for every
output. In particular, in this case, step 2 increases the upper bound
on the size of the BDDs by an extra factor of 8 [Eq.~\eqref{eq:step2}
combines a function of many variables with two functions of the same 3
variables; the factor of 8 comes from the worst case branching due to
the 3 variables]. The sizes would increase to
$\sim (7\times 8)^\ell = n^{\log_3(7\times 8)}$.

\end{proof}

We then proceed to extend the bound on the BDD sizes, obtained for the
1-bit NOT gate, to the case where the chip is seeded by a CNOT or a
Toffoli gate. We start by considering a situation in which all the
bitlines acted on by a CNOT or a Toffoli gate overlap with a single
gate of the first layer of $\hat N$. In this case, the conjugation by
that gate would result in a 3-variable BDD, no different in structure
from those encountered above in the context of the conjugation of the
NOT gate by the first layer of $\hat N$. In this case, the scaling of
the BDDs following conjugation with the subsequent layers will proceed
analogously, as these layers will bring fresh variables, leading to
the same scaling of the BDD size with the number of layers as
above. On the other hand, the case in which the bitlines acted on by
the CNOT or Toffoli gate overlap with a gate in more distant layers of
the conjugation process, the scaling becomes less favorable (albeit
still polynomial). Instead of going through the more complex
derivation of bounds in this context, we explore a simplification
lended by considering a system with two registers, a configuration
which, in any case, is needed if one is to operate on (e.g. add or
multiply) two data values.


\subsubsection*{Two registers}:
\label{sec:two-registers}

\vspace{0.2cm}

\begin{figure}[h]
  \centering
  \vspace{0.5cm}
  \includegraphics[angle=0,scale=0.3]{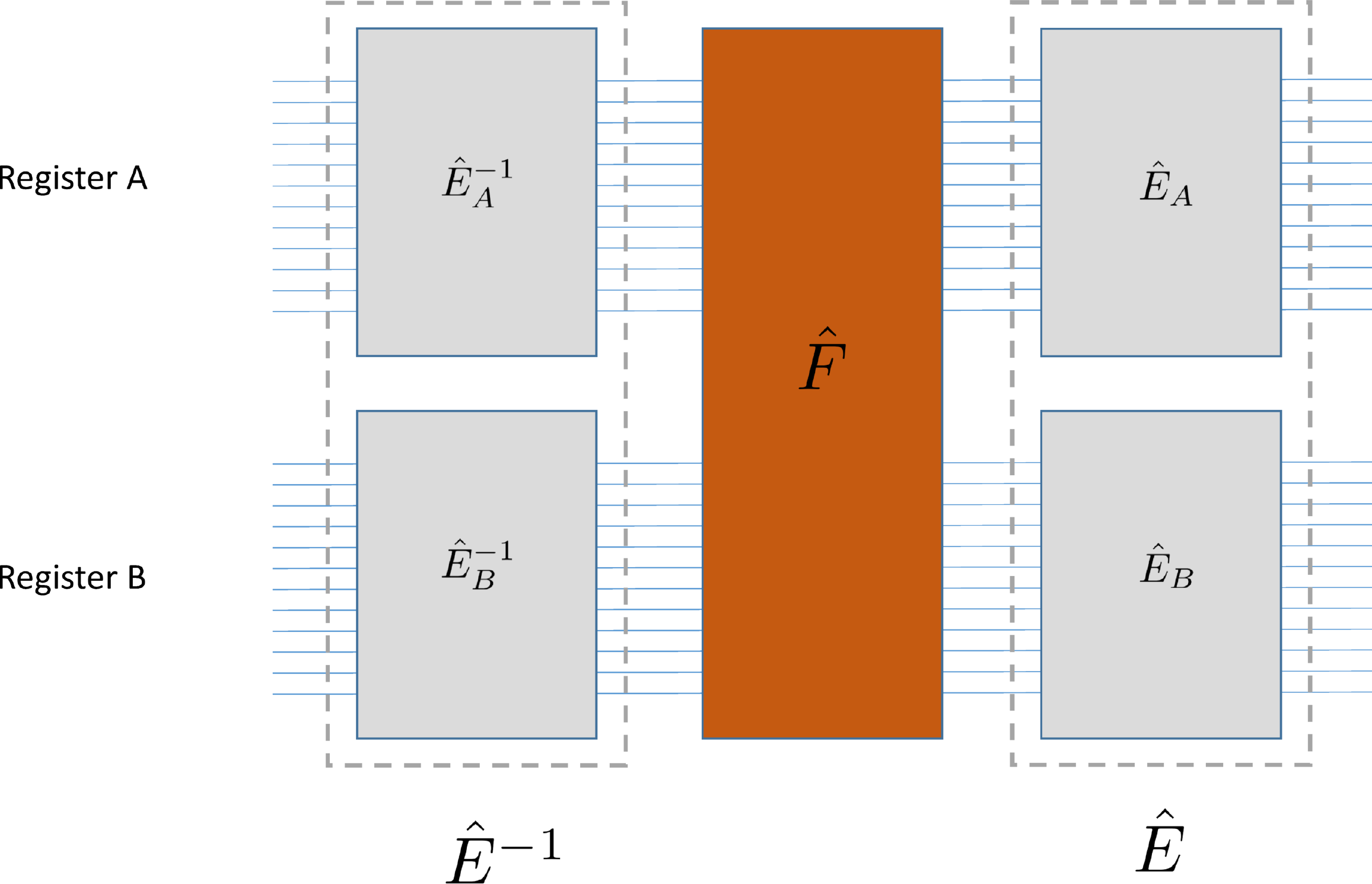}
  \vspace{0.5cm}
  \caption{Conjugation of a function $F$ (operator $\hat F$) that acts
    on data contained in two separate registers, $A$ and $B$,
    encrypted with encrypted with reversible circuits $E_A$ and $E_B$,
    each of which is structured as a two-stage cipher as shown in
    Fig.~\ref{fig:register_cipher}.}
  \label{fig:two_registers}
\end{figure}

If there are two registers, one replaces the encryption and decryption
operators $\hat E$ and ${\hat E}^{-1}$ with tensor products of
operators acting independently on each of the registers.
Specifically, two registers $A$ and $B$, containing data $x_A$ and
$x_B$ are encrypted with reversible circuits $E_A$ and $E_B$,
respectively. (We note that circuits $E_A$ and $E_B$ can be the same
or different.) In this construction, the operator $\hat E$ in
Eq.~\eqref{eq:F_E} is replaced by the two-register operator
$\hat E \equiv \hat E_A \otimes \hat E_B$, with the operator $\hat F$
representing the function $F$, which acts on both $x_A$ and $x_B$, now
bridging across the two registers, as illustrated in
Fig.~\ref{fig:two_registers}.

Let us first consider the case of a CNOT gate in a two-register
configuration. When the target and control bitlines land in different
registers, the line of argument follows the discussion of the NOTs
above, with the same scaling of the size of the BDDs with the size of
the individual registers. When the target and control act on bitlines
in the same register, we can use SWAP gates to move either the control
or target to the other register. In this case, we increase the number
of gates to be conjugated by a factor of 3 -- the original CNOT is
replaced by three gates, two SWAPs and a CNOT, all bridging two
registers. We thus increase the number of chips three-fold, but reduce
the conjugation problem to one similar to that already encountered for
NOT gates.

In the case of a Toffoli gate, regardless of the position of the
bitlines on which the gate acts, one can use SWAPs to move bitlines so
that two of them overlap with a gate in the first layer of $\hat N$ in
one register, with the third bitline located in the other
register. The cost of this move is the addition of at most four SWAPs,
thus increasing the number of chips by a factor of 5. Again, with this
movement of bitlines via SWAPs we reduce the conjugation problem to
one similar to that already considered above without changing the
scaling of the size of the BDDs associated with individual chips.

The above arguments, together with Lemma \ref{lemma:NOT}, establish
that chip BDDs are polynomial-sized, as stated in Theorem~\ref{thm:2}.

\section{Assumption~\ref{A2}: Best Possible Obfuscation with injection of randomness}
\label{sec:security}

The security of the EOC scheme hinges on whether or not one can
recover significant information about $\hat E=\hat N\hat L$ and
$\hat F$, given access to the collection of chips that represents the
encrypted function $\hat F^E$. We address this question in two
steps: the first concerns the security of a single chip in isolation,
and the second the security of the entire collection of chips.

\subsection{Single chip}
\label{sec:single-chip}

As mentioned above, our process of building the chip via conjugation
with the cipher $\hat E$ implements Best Possible Obfuscation,
introduced by Goldwasser and Rothblum~\cite{Goldwasswer-Rothblum}.
Their paper considered a class of functions that are computable by
POBDDs, and showed that normal-form POBDDs are themselves the
best-possible obfuscators of those functions. For that class, the
best-possible obfuscator (which outputs the POBDD) is also an
indistinguishability obfuscator~\cite{Barak}. The critical question
addressed in this section is whether, in the context of a circuit
built via conjugation, Best Possible Obfuscation is ``Good Enough'',
i.e., whether the chip POBDDs hides the nonlinear gates in $\hat N$
and the initial gate being conjugated.

\begin{figure}[h]
  \centering
  \vspace{0.5cm}
  \includegraphics[angle=0,scale=0.90]{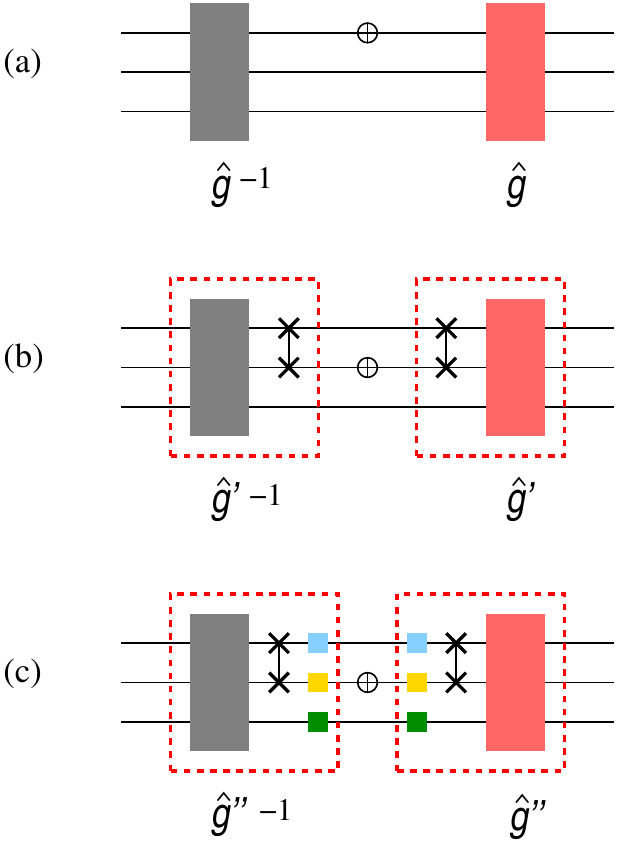}
  \vspace{0.5cm}
  \caption{Equivalent factorizations of (a)
    $\hat h=\hat g\;\sigma_1^x \;\hat g^{-1}$, where $\sigma_1^x$
    corresponds to a NOT gate placed on bitline 1. (b) The NOT gate
    can be moved, for example, to bitline 2 by absorbing SWAP gates
    between lines 1 and 2 into the right of $\hat g^{-1}$ and into the
    left of $\hat g$, yielding the factorization
    $\hat h=\hat g'\;\sigma_2^x \;\hat {g'}^{-1}$. (c) Independent
    pairs of either identity or NOT gates (shown in matching colors)
    can be inserted and then absorbed to the left and right,
    respectively, into the gates $\hat g'^{-1}$ and $\hat g'$ of (b),
    yielding the factorization
    $\hat h=\hat g''\;\sigma_2^x \;\hat {g''}^{-1}$.}
  \label{fig:small-chip}
\end{figure}

We proceed by considering a simple example of a small chip built via
conjugation of a single NOT gate with only one layer of nonlinear
gates, as depicted in Fig.~\ref{fig:small-chip}a. The NOT gate can be
viewed as a chip at level $\ell=0$, and the result of conjugation is a
3-bit chip at level $\ell=1$, written in operator form as
$\hat h=\hat g\;\sigma_1^x \;\hat g^{-1}$, where $\sigma_1^x$ is the
NOT operator that flips the value of bit 1, $x_1\to \bar x_1$. The
functionality of the 3-bit chip is encoded in three BDDs,
$h_{i}(x_1,x_2,x_3), i=1,2,3$, describing the outputs of the gate
operator $\hat h$. While the BDDs are unique normal forms, there are
multiple ways of factoring the operator $\hat h$. Examples are
illustrated in Fig.~\ref{fig:small-chip}. For instance, in
Fig.~\ref{fig:small-chip}b we inserted identities in terms of SWAP
gates exchanging bitlines $1$ and $2$ (alternatively, we could have 
exchanged bitlines $1$ and $3$). In Fig.~\ref{fig:small-chip}c, we also
randomly inserted pairs of NOT gates on each of the three bit lines,
and absorb them into a redefinition of the 3-bit gate and its
inverse. (One may go further and introduce a random reversible 2-bit
gate in $S_4$ and its inverse acting on bits 2 and 3, and absorb these
gates onto the 3-bit gates.)  These simple examples illustrate
multiple factorizations of $\hat h$ and reflect ambiguities in its
factors: the initial NOT being conjugated could have been in bitline
$i=2,3$ instead of 1, and the cipher gate could have been $\hat {g''}$
instead of $\hat g$, with any choice of placement of NOTs on any of
the three bitlines. Meanwhile, BDDs do not distinguish particular
factorizations, as they only encode the functionality of the product,
$\hat h$.

One can generalize the example above to chips build after arbitrary
layers of conjugation. In particular, chips at level $\ell+1$: (i) do
not contain sufficient information to determine which bitlines the
chip at level $\ell$ acted on; and (ii) cannot distinguish among the
multiple choices of bit negations, inserted in pairs on any of the
fresh bitlines (those not contained in the chip at level $\ell$). Due
to these ambiguities, it is impossible, by examining only the BDDs, to
reverse the hierarchical structure that determines how bits are
grouped in triplets for the placement of the 3-bit gates of the cipher
and the exact 3-bit gates that were deployed. Basically, information
is erased (not only concealed) in the process of assembling a {\it
  single} chip via conjugation.

\begin{figure}[h]
  \centering
  \vspace{0.5cm}
  \includegraphics[angle=0,scale=0.47]{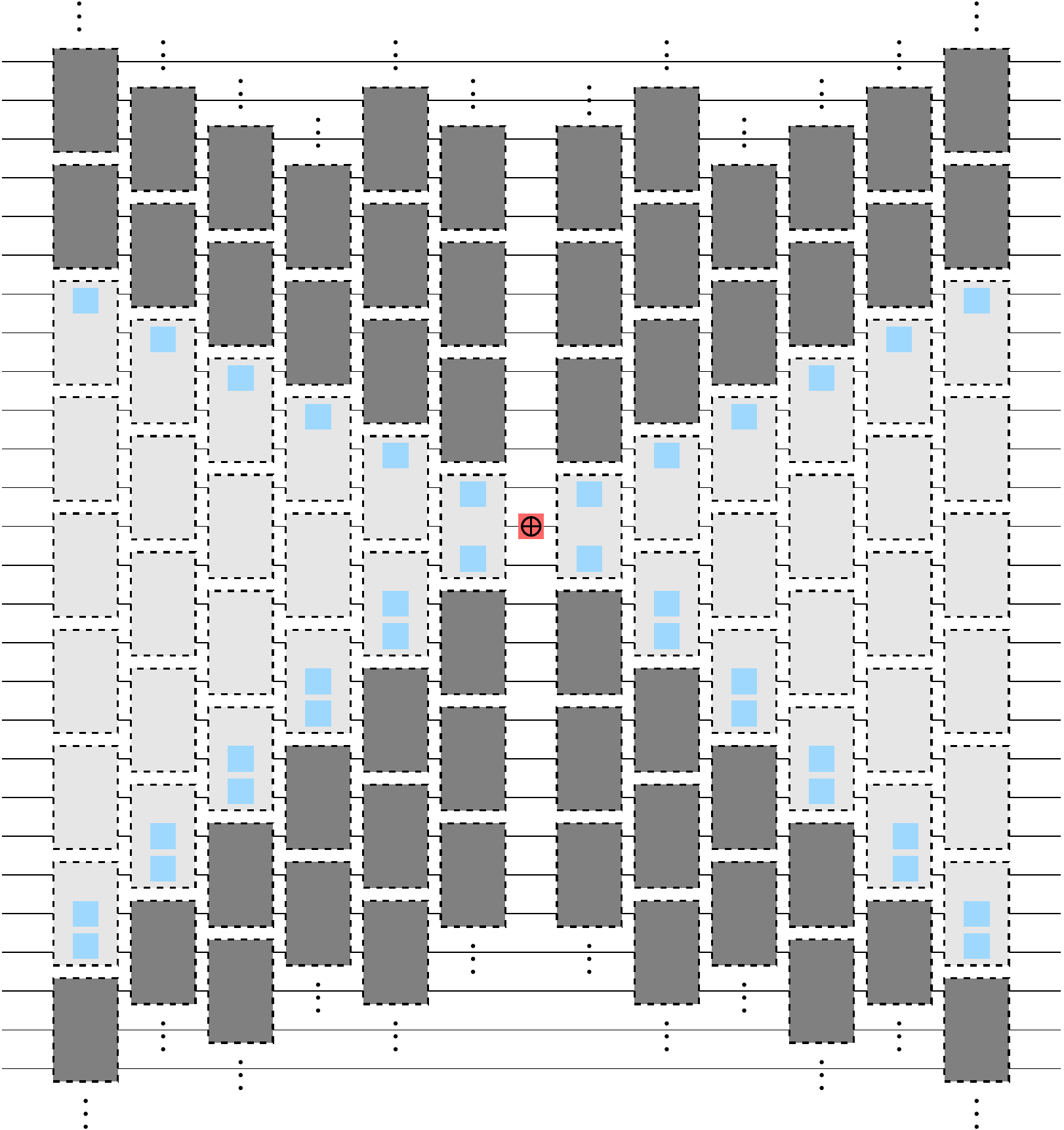}
  \vspace{0.5cm}
  \caption{Region of influence of a 1-bit NOT gate (red square in the
    middle) in the course of conjugation. The circuits to the left and
    right of the NOT gate represent $\hat N^{-1}$ and $\hat N$,
    respectively. For simplicity, the figure displays a
    one-dimensional ($D=1$) ``brickwall'' arrangement of the fully
    packed circuit of 3-bit gates. (The description generalizes to all
    dimensions $D$, including the tree-like structure corresponding to
    $D\to\infty$.) Each 3-bit gate is depicted as a rectangle, with
    the horizontal (``time'') direction represents the order of the
    computation. The 3-bit gates depicted as dark rectangles -- the
    collection of which defines the ``dark region'' -- are not
    affected by the NOT gate in the middle and annihilate pairwise (a
    $g^{-1}$ from $\hat N^{-1}$ and a $g$ from $\hat N$). Thus these
    gates are ``invisible'' and do not enter in the composition of the
    chip. By contrast, the 3-bit gates depicted as light rectangles
    {\it do} contribute to the buildup of the chip and are contained
    within the ``past'' and ``future'' light cones (shown as light
    blue squares) to the left and right of the NOT, respectively.}
  \label{fig:lightcone}
\end{figure}

We also note that, in the course of conjugation, gates in the
$\ell+1$-th layer of $\hat N$ that do not touch the bits of the chip
at level $\ell$ simply annihilate in pairs, and thus information on
those gates is erased. Trivially, the BDDs are insensitive to these
gates. The information lost has a simple interpretation in the context
of operator spreading in physical systems (see, for example,
Ref.~\cite{Roberts2015}): it corresponds to the dark region outside of
the light cone of influence that develops through the unitary
evolution of an operator, in our case the initial gate being
conjugated. This notion is illustrated in Fig.~\ref{fig:lightcone},
where we depict, for simplicity, a one-dimensional ($D=1$) version of
the tree-like ($D\to \infty$) packing of 3-bit gates in each layer.

The arguments given above reinforce the fact that, at the level of a
single chip, Best Possible Obfuscation is indeed "Good Enough" in
removing access to the cipher and in hiding the gate being
conjugated. What is not a priori obvious is what level of security
Best Possible Obfuscation of individual chips confers to the
concatenation of chips representing the full function $\hat F^E$. Can
correlations extracted from the full collection of BDDs describing
multiple chips reveal any details about the cipher?

\subsection{Multiple chips: injecting randomness}

\begin{figure}[!h]
  \centering
  \vspace{0.5cm}
  \includegraphics[angle=0,scale=0.8]{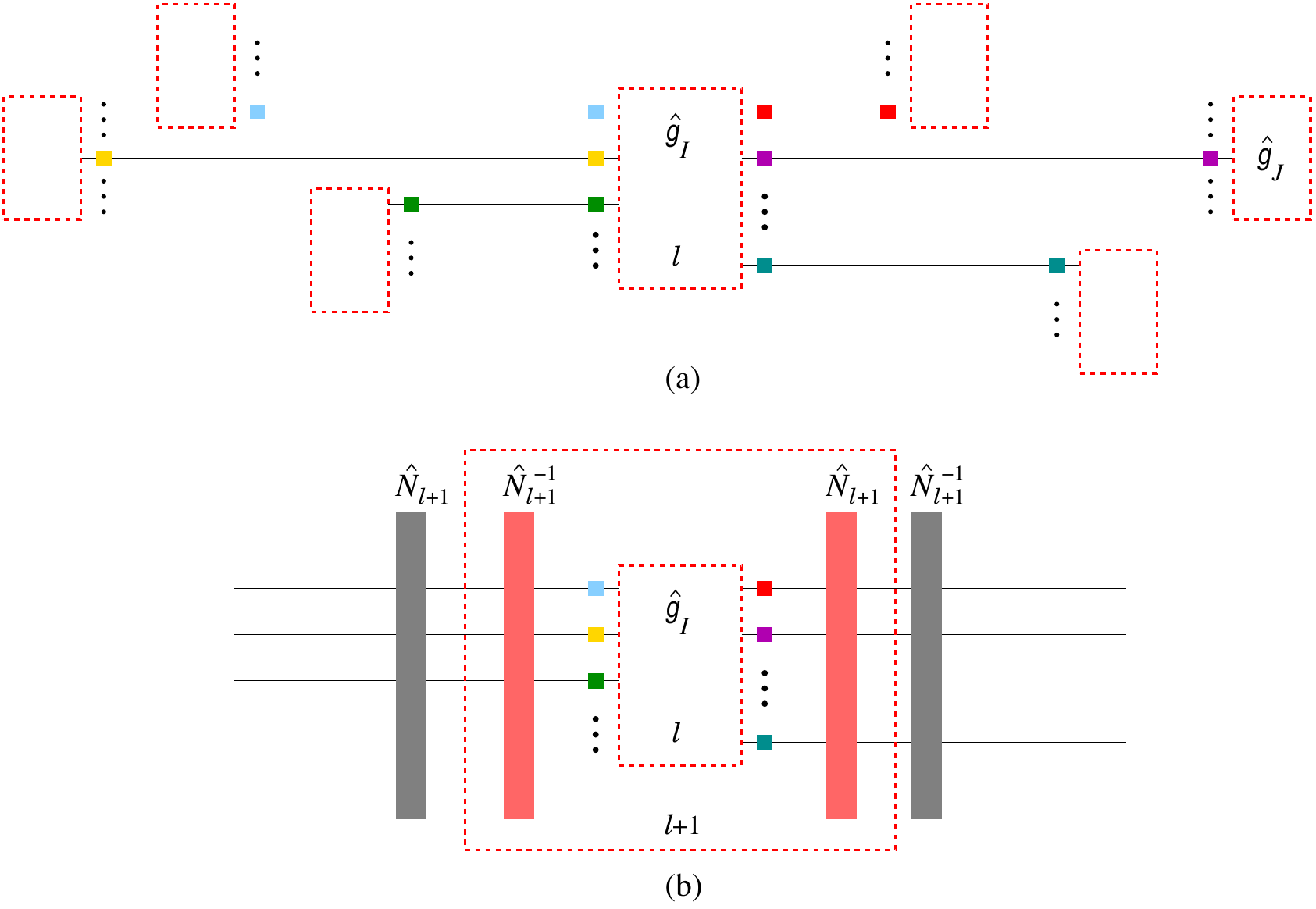}
  \vspace{0.5cm}
  \caption{Construction of the chips that incorporate the injection of
    randomness. The construction is carried out recursively,
    layer-by-layer of $\hat N$ as follows: (a) we randomly insert
    pairs of either identity operators or NOTs (depicted with matching
    colors) on internal wires connecting two level-$\ell$ chips, $I$
    and $J$; we then absorb one operator into the output of the chip
    on the left and the other into the input of the chip on the right;
    and (b) we conjugate by $\hat N_{\ell+1}$, the $\ell+1$-th of
    $\hat N$, arriving at the chip at level $\ell+1$, represented by
    the larger dashed box.}
  \label{fig:edge}
\end{figure}

Instead of answering this question, we obviate it by incorporating
randomness, which washes out correlations among chips by scrambling
the functionality of individual chips while preserving the
functionality of the full computation.  Below we present the
construction of a new set of chips, ${\hat g}^{N,\eta}_{i,q}$, that
incorporates randomness, the presence of which is symbolized by
$\eta$. For notational simplicity we group the subscripts $i,q$ into a
super index $I$, and concentrate on the construction of the chip by
that label, i.e., that initiated by ${\hat g}_I$.  All chips at level
$\ell+1$ are built recursively (and in parallel) from the chips at
level $\ell$, according to the following three-step process: (1)
randomly insert pairs of either identity operators or NOTs on internal
wires connecting two level-$\ell$ chips, $I$ and $J$ (see
Fig.~\ref{fig:edge}a); (2) on the edges that received a pair of
operators, absorb one operator into the output of the chip on the left
and the other into the input of the chip on the right; and (3) proceed
with conjugation by layer $\ell+1$ of $\hat N$ (see
Fig.~\ref{fig:edge}b), in the exact same manner described in
Sec.~\ref{sec:nonlinear}, this time in synchrony for all gates that
emerged from the linear stage of the cipher that, together, represent
the full function $\hat F^E$.


This process illustrated in Fig.~\ref{fig:edge} randomizes the
functionality of individual chips while preserving the functionality
of $\hat F^E$. We also stress that, in general, the NOT gates that are
injected into the chips do not commute with the nonlinear gates in
$\hat N$ and thus, the scrambling effect of the NOTs is nonlinearly
amplified through the conjugation process.

Randomization induced via absorbtion of NOTs in the second step above
is trivially reflected in the BDDs of all resulting chips. For every
input of a chip that incorporates a NOT, one flips the decision
branches of the corresponding nodes (with that input variable) of the
BDDs, i.e., a solid branch (or true) is switched to a dashed branch
(or false), and {\it vice-versa}. Similarly, for every output of the
chip that incorporates a NOT, one swaps the $\top$ and $\bot$ terminal
nodes. Notice that the BDD retains its size, as no new nodes are
created by the randomization process. The third step of the process,
the conjugation with the next layer, proceeds exactly as before (see
Sec.~\ref{sec:nonlinear}) and thus, the bounds on the scaling of the
sizes of BDDs obtained in Sec.~\ref{sec:scalingBDD} remain unchanged.

Another way of evaluating the effect of the randomization process is
to consider in more detail the ambiguity it adds to a chip beyond that
already present in the absence of randomness (i.e., that already
accounted for in our discussion of the single chip above).  Recall
that, in the absence of randomness, we argued that there was ambiguity
in the choice of negations in 2/3 of the bitlines of the chip,
corresponding to the fresh bits present in the chip at level $\ell+1$
but absent in the chip at level $\ell$. In the presence of the added
randomness, the ambiguity is extended to every bitline, including the
1/3 of the lines inherited from the chip at the previous level. The
BDDs are oblivious to 2/3 of negations because they are normal forms,
and scrambled by the other 1/3 of negations.

While it is difficult to quantify the amount of information that an
adversary can extract about the cipher from the full collection of
chips representing $\hat F^E$, the arguments above provide a lower
bound on what {\it cannot} be extracted: one cannot resolve whether
bitlines are negated or not in between the layers of $\hat N$. In
other words, one cannot resolve between a layer $\hat N_\ell$ and
another putative layer $\hat {\tilde N}_\ell=\hat N_\ell\;\hat P_\ell$, where
$\hat P_\ell$ is a layer of random NOTs. There are $2^{n\log_3 n}$
possibilities for incorporating negations before all layers, so
building $\hat F^E$ increases entropy by, at the very least,
$\Delta S={n\log_3 n}$. It is possible that this increase in entropy
is sufficient to guarantee that one cannot break the encryption. The
situation parallels that of the Advanced Encryption Standard (AES),
where all information about the cipher is exposed except for the
bitwise XORs in the AddRoundKey iterations. In the case of AES, the
negations are deterministically generated round-by-round from the
key. While the parallel to AES is not a proof of security (and AES has
not yet been formally proved secure either), we note that the entropy
generated in EOC is superextensive in $n$, while that in AES scales
linearly with the size of the key. We further note that the above
estimate of the entropy is a very conservative lower bound to the
information lost in assembling the chips, as it does not account for
the lack of knowledge about, for example,
(1) the specific gates in $S_8$ that are used, beyond
the effects of negations above; and (2) the connectivity of the gates
in the cipher, i.e., the triplets of bits on which each of the 3-bit
gates act on.

We conclude that inserting randomness confers security to EOC beyond
that provided by Best Possible Obfuscation of individual chips in the
absence of disorder. The discussion and arguments presented in this
section are the basis for our Assumption~\ref{A2}. The security of the
EOC scheme hinges on this assumption, along with Assumption~\ref{A1}.


\section{Conclusions and open directions}
\label{sec:conclusions}

This paper introduces EOC as an alternative to currently used schemes
for Fully Homomorphic Encryption. EOC utilizes reversible logic, and
can be viewed intuitively as way to perform computation of an operator
(function) $\hat F$ in a transformed frame (defined via a unitary
transformation, $\hat F^E = \hat E \hat F \hat E ^{-1}$) in which the
transformed operator $\hat F^E$ (the encrypted program) acts on
transformed state vectors $\hat E \ket{x}=\ket{E(x)}$ (the encrypted
data). Throughout, we refer to the obfuscated implementation of the
operator $\hat F^E$ as the evaluator $O(\hat F^E)$. The obfuscation of
$\hat F^E$ relies {\it specifically} on the conjugated form of this
operator and on the structure of the cipher $\hat E$.

The evaluator is expressed as a sequence of chips,
with the $n$ outputs of each chip expressed as BDDs. We placed exact
polynomial upper bounds on the number of chips and on the size of the
chip-BDDs. Explicitly, we proved that: (1) the number of chips, for
each elementary gate in $\hat F$, is bounded by $n^{\mu_3}$, with
${\mu_3}=3\;\log_2 3 \approx 4.75$; and (2)the size of the BDDs for
each output of the chip has at most $n^{\gamma}$ nodes, with
$\gamma=\log_3 7\approx 1.77$. It follows that the time complexity of
EOC (per gate in $\hat F$) is bounded by $n^{\mu_3+2}$ or
$n^{\mu_3+1}$ if the chip BDDs are evaluated in series or in parallel,
respectively; and that storage space required, as measured by the
number of nodes in all BDDs, is bounded by $n^{\mu_3+\gamma+1}$.

Establishing these bounds hinges on the $\OO(\log n)$ depth of both
linear and non-linear stages of the cipher, the security of which was
posited in Assumption~\ref{A1} in Sec.~\ref{sec:contribution} and
motivated in Sec.~\ref{sec:cipher}. The discussion of the 2-stage
cipher in Sec.~\ref{sec:cipher} connects security against differential
attacks to measures used to diagnose chaos and irreversibility in
quantum circuits developed in Ref.~\cite{cipher-paper}.

Having established the polynomial complexity of EOC with $n$, we
turned to the question of its security. We presented different
elements of the EOC construction in Sec.~\ref{sec:contribution},
Sec.~\ref{sec:evaluator} and Sec.~\ref{sec:security} ,which provide
the basis for our Assumption~\ref{A2} on the security of the approach,
namely that EOC yields an Indistinguishability Obfuscation of the
conjugated circuit $\hat F^E$. There are three contributing mechanisms
to the loss of information that, taken together, support this
assumption. The first is based on the unitary-transformation form of
conjugation (which contains both $\hat E$ and $\hat E^{-1}$), and
involves complete erasure of information: gates that do not touch the
footprint of the chip at a particular level of conjugation annihilate
in pairs and thus, as conjugation proceeds layer by layer, gates of
the cipher outside of the ``light cone'' are simply invisible. The
second mechanism is the compression of information enabled through the
representation of chip outputs as collections of polynomial-sized
BDDs, which are normal forms that only expose the minimum information
required to establish functionality. As discussed in the body of the
paper, a BDD-based chip realizes the notion of Best Possible
Obfuscation introduced by Goldwasser and
Rothblum,~\cite{Goldwasswer-Rothblum} which, in turn, we argue is
sufficient for the security of individual chips. However, the level of
security that Best Possible Obfuscation implies for the concatenation
of chips representing the result of the full computation of $\hat F^E$
is not a priori obvious. We circumvented this issue by incorporating a
third (external) source of obfuscation; namely, before each level of
conjugation by a nonlinear layer, we add random pairs of NOTs
(identities) between chips, which are then separated and incorporated
into inputs and outputs of chips connected by a given bitline. This
randomization process scrambles the functionality of individual chips
while preserving the functionality of the overall conjugated
circuit. This injection of randomness across the system washes out
correlations among chips that might have revealed information on the
cipher $E$ or function $F$, and thus confers security for the fully
conjugated circuit beyond that provided by Best Possible Obfuscation
of individual chips.

Layers of $\OO(n)$ random negations, interspersed between
$\OO(\log n)$ layers of nonlinear gates, inject entropy of order
$\OO(n\;\log n)$. We note that the effect of adding NOTs between
layers of nonlinear gates is similar to that in the AddRoundKey rounds
of AES. How to precisely quantify the total information, either
invisible (i.e., truly erased) or just obfuscated through the
combination of nonlinear conjugation that amplifies the effect of
insertions of NOTs at random, remains an open question. Future work is
needed in order to develop a formal framework for quantifying the
actual entropy of conjugated circuits and their associated BDDs,
analogous to the statistical mechanics approach to the security of
shallow $\log n$-depth ciphers based on string entropy and
out-of-time-ordered correlators (OTOCs) described in
Ref.~\cite{cipher-paper}. It is appealing to speculate that the
insights on random classical circuits built on the mapping to string
space can be extended to the study of more general reversible circuits
and their compression via BDDs.

We remark that, for practical purposes, the flexibility afforded us by
the random padding of the input (enabling probabilistic encryption)
can be used to greatly decrease the overhead of the EOC scheme, if
combined with countermeasures against ciphertext attacks that are
enabled by the presence of ancilla bits. The significant simplification
amounts to replacing the linear stage $\hat L$ of
Fig.~\ref{fig:register_cipher} with a specially designed single layer
of inflationary gates. In this case, one can deploy the conjugation
scheme with the BDDs including this one linear layer (along with
$\log_3 n-1$ nonlinear layers). Correspondingly, the polynomial
complexity drops by a factor of $n^{\mu_3}$. In particular the
execution-time overhead becomes $n^{2}$ or $n^{1}$ if the chip BDDs
are evaluated in series or in parallel, respectively. We hope to
explore this possibility in a future publication.

In closing, we stress that the EOC framework -- based directly on
logic elements as building blocks -- makes a hardware implementation
natural. For example, by implementing EOC in silicon (e.g., in field
programmable gate arrays -- FPGAs; or Application-specific Integrated
Circuits -- ASICs) one could reach speeds of computation on encrypted data
limited only by state-of-the-art electronics. Such a practical
deployment of EOC would impact all aspects of data science for which
security and privacy are essential.

\section*{Acknowledgments}

The authors would like to thank Ran Canetti and Adam Smith for many
enlightening conversations and continuous encouragement, and for
stimulating us to plunge into these problems with a physics-inspired
approach. We are especially grateful to Adam Smith for detailed comments and suggestions on a previous version of the manuscript.

{\bf Statement of work:} E.R.M. contributed to
Sec.~\ref{sec:linear}; J.J.-S. contributed to
Secs.~\ref{sec:nonlinear}; and C.C. and A.E.R. contributed to all sections
of the paper.

\appendix

\setlength{\parskip}{5mm plus3mm minus3mm}
\counterwithin{figure}{section}
\section{Reversible logic equivalences: collisions, simplifications,
  and factorizations}
\label{sec:tools}

Any reversible circuit associated with an even permutation can be
broken down into elementary NOT, CNOT, and Toffoli controlled gates
(odd permutations require an additional ancilla bit). The NOT gate
negates bit $j$ irrespective of all others: $x_j\to \bar x_j =
x_j\oplus 1$. The CNOT gate negates (the ``target'') bit $j$
conditional on whether another (``control'') bit $i$ is 0 or 1:
$x_j\to x_j\oplus x_i$. Finally, the Toffoli gate negates (the
"target") bit $j$ conditional on whether ("control") bits $i_1$ and
$i_2$ are both true: $x_j\to x_j\oplus x_{i_1}\,x_{i_2}$.

More generally, one can define controlled gates with controls over $n$
bit lines. (These gates can be broken down into smaller gates with
$n<2$ using the factorization rules below.) Consider a Boolean
expression $B(x_{i_1}, \dots, x_{i_n})$ that depends on the logic
variables $x_{i_1}, \dots, x_{i_n}$ in bit lines labeled by
$i_1,\dots,i_n$.  A generic control gate should negate a target bit
$j\ne i_1,\dots,i_n$, i.e. $x_j \to \bar x_j$, if $B$ is {\tt True}
($B=1$), and leave $x_j$ untouched if $B$ is {\tt False} ($B=0$). In
other words, a target bit, $j$, that is acted on by this gate will be
modified as follows:
\begin{align}
  x_j \to x_j \oplus B(x_{i_1}, \dots, x_{i_n})
  \;,
  \quad
  j\ne i_1,\dots,i_n
  \;.
\end{align}
We shall concentrate on Boolean expressions that can be expressed as a
product involving either the variables $x_{i_k}$ or their negation
$\bar x_{i_k}$,
\begin{align}
  B(x_{i_1}, \dots, x_{i_n})
  =
  \left(x_{i_1}\oplus \sigma_{1}\right)
  \;
  \left(x_{i_2}\oplus \sigma_{2}\right)
  \;
  \cdots
  \left(x_{i_n}\oplus \sigma_{n}\right)
  \;,
\end{align}
where the variables $\sigma_{k}=0,1$, for $k = 1,\dots,n$, determine
the polarities of the $n$ controls entering the expression for $B$:
$x_k\oplus 0 = x_k$, and $x_k\oplus 1 = \bar x_k$.

To describe a controlled gate, $g$, we need the following information:
the bit line corresponding to the target bit $t(g)$ and the Boolean
expression acting on the control bits, which is cast as a set $C(g)$
of pairs with the control bits and their polarities. In the example
above, $t(g)=j$ and $C(g)=\{(i_1,\sigma_1),\dots,(i_n,\sigma_n)\}$. We
denote by $b(C)$ the set of only the bits $\{i_1,\dots,i_n\}$, and by
$|C|$ the number of control bits in $C(g)$ ($|C|=n$ in the
example). We also refer to the target bit $t(g)$ as the ``head'' of
the gate and the set $C(g)$ as its ``tail'', which contains all the
information on the controls, i.e., the control bits and their polarity
for the controlled-Boolean expression.

The NOT, CNOT, and Toffoli gates are particular examples of
controlled-Boolean gates. The NOT corresponds to a constant $B=1$,
which depends on no other bits ($C=\varnothing, |C|=0$).  The CNOT has
$|C|=1$, and corresponds to one of two possible Boolean functions,
$B=x_{i_1}$ or $B=\bar x_{i_1}$, associated to positive or negative
polarity controls, respectively. The Toffoli gate has $|C|=2$, and
corresponds to one of four possible Boolean functions:
$B=x_{i_1}\,x_{i_2}$, $B=\bar x_{i_1}\,x_{i_2}$, $B=x_{i_1}\,\bar
x_{i_2}$, or $B=\bar x_{i_1}\,\bar x_{i_2}$ depending on the choice of
polarities of the controls. An example of a gate with $|C|=3$ controls
is one with $B=x_{i_1}\,x_{i_2}\,\bar x_{i_3}$.

\begin{figure}[h]
  \centering
  \includegraphics[angle=0,scale=0.95]{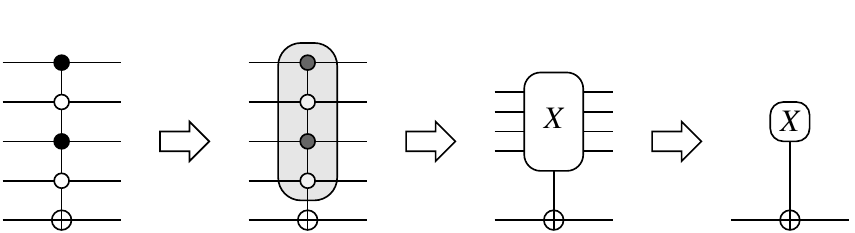}
  \caption{A simplified representation of a Boolean control gate. The
    target bit is marked by a $\oplus$. The control (tail) bits are
    lumped into a box, here represented by $X$. Notice that control
    bits inside the box can be negated (white circle) or not (black
    circle). The bit lines going in and out of the box are omitted.}
  \label{fig:gate_representation}
\end{figure}

Here it is useful to introduce the simplified graphical representation
of a control gate illustrated in
Fig. \ref{fig:gate_representation}. The Boolean expression
representing controls (tail) is lumped into a single control box, $X$,
which can cover multiple bit lines, that are not necessarily
consecutive. While we often omit control bit lines that go in and out
of the control box, when particular control bits play a significant
role (e.g., in collisions, see below) they are pulled out of the box
and shown explicitly.

We next introduce equivalence rules for the action of multiple
Boolean-controlled gates which reflect the non-commutativity of the
actions of sequential gates.  (Similar rules have been discussed in
the context of circuit simplification, e.g., Ref.~\cite{Iwama2002}.)
We refer to these circuit equivalences as ``collision'' rules, as they
express the fact that interchanging the order of two non-commuting
controlled gates $\hat g$ and $\hat h$ requires inserting ``debris''
gates in between.  More precisely, when $g$ and $h$ do not commute,
i.e., $\hat g\;\hat h\ne \hat h\;\hat g$, then $\hat g\;\hat h = \hat
h\;\hat D\;\hat g$, in which case the "debris" $\hat D= \hat
h^{-1}\;\hat g\;\hat h\;\hat g^{-1}$ can also be broken down in terms
of controlled gates. (In the particular case when $g$ and $h$ commute,
$D=\openone$.)

  
\subsection{Collision rules for controlled gates}
\label{sec:collision_rules}

\begin{figure}[h]
  \centering
  \includegraphics[angle=0,scale=0.75]{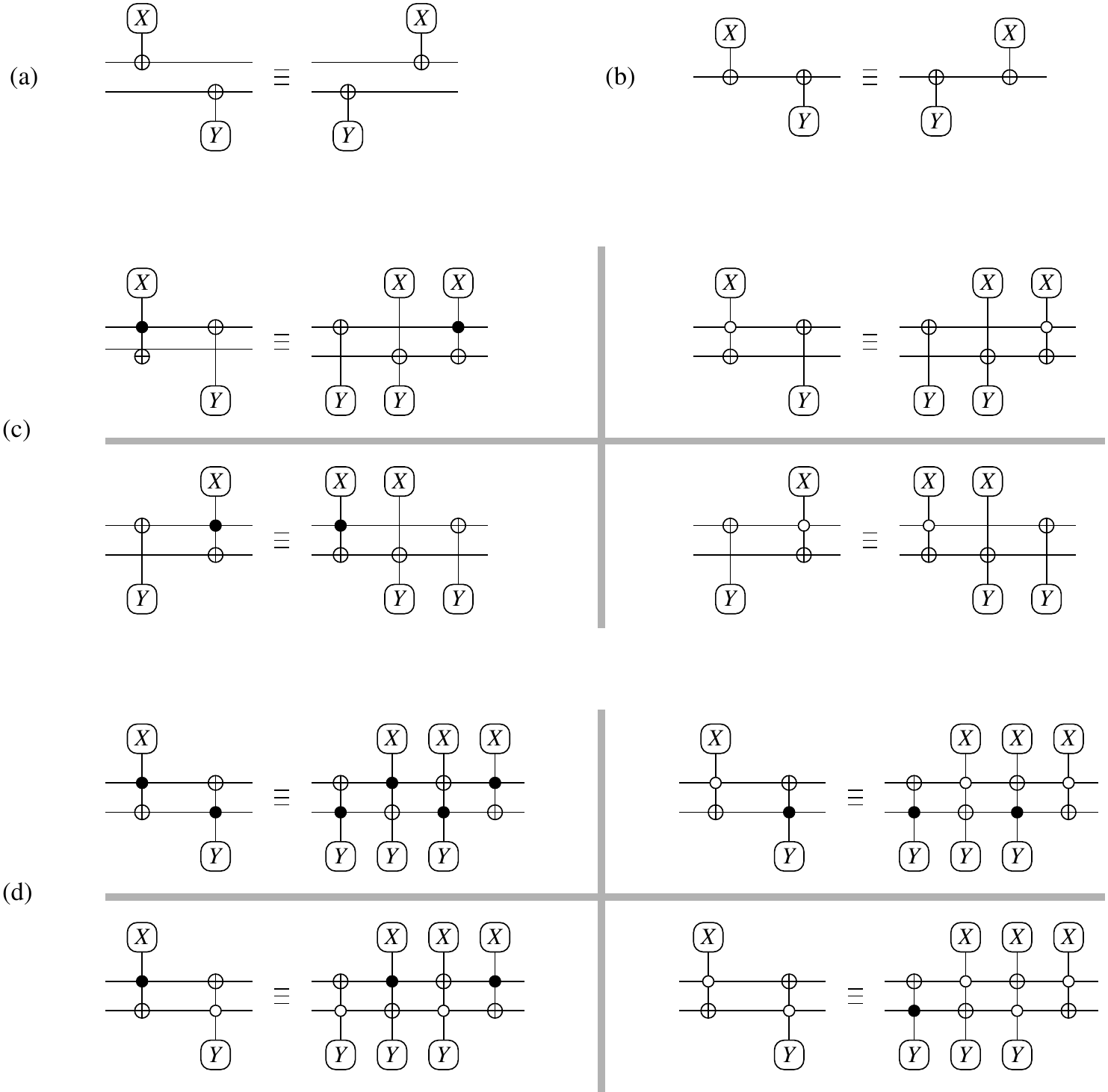}
  \caption{Collision rules. (a) No-collision rule: The head of one
    gate does not share a bit line with any of the bits of the other
    gate, and vice versa. Notice that tails $X$ and $Y$ can overlap
    and run over some common bit lines.
    (b) Head-on-head collision rule: The heads share the same bit
    line. Notice that the tails $X$ and $Y$ can overlap.
    (c) One-head collision rules for four gate types and
    configurations: The head of one gate shares a bit line with a bit
    in the tail of the other gate, but not vice versa. The gates with
    both $X$ and $Y$ correspond to having the product $XY$ as
    control. The tails $X$ and $Y$ can overlap.
    (d) Two-head collision rules for four gate types and
    configurations: The head of one gate shares a bit line with a bit
    of the tail of the other gate, and vice versa. Again, the gates
    with both $X$ and $Y$ correspond to having the product $XY$ as
    control, and the tails $X$ and $Y$ can overlap.
  }
  \label{fig:collisions_combined}
\end{figure}

The collision rules are illustrated graphically in
Fig.~\ref{fig:collisions_combined}.  The simplest examples, of
commuting gates, $\hat g$ and $\hat h$, are shown in
Figs.~\ref{fig:collisions_combined}(a) and
~\ref{fig:collisions_combined}(b).  These represent cases in which (a)
there is no overlap between the target bit of one gate and the control
bits of the other, and vice-versa; or (b) the gates share the same
target bit (coinciding heads). Hereafter we refer to these cases as
{\it no collision} and {\it head-on-head collision}, respectively.  A
non-trivial collision occurs when the target line of one gate overlaps
with control lines of the other. Clearly, in this situation the gates
do not commute, and interchanging their order requires the insertion
of additional "debris" gates in order to preserve the functionality of
the original order.  Figs.~\ref{fig:collisions_combined}(c) and
~\ref{fig:collisions_combined}(d) illustrate non-trivial collisions
cases, which we refer to as {\it one-headed} and {\it two-headed}
collisions, respectively.



\subsection{Simplification and polarity mutation rules}
\label{sec:simplification}

\begin{figure}[h]
  \centering
  \includegraphics[angle=0,scale=0.75]{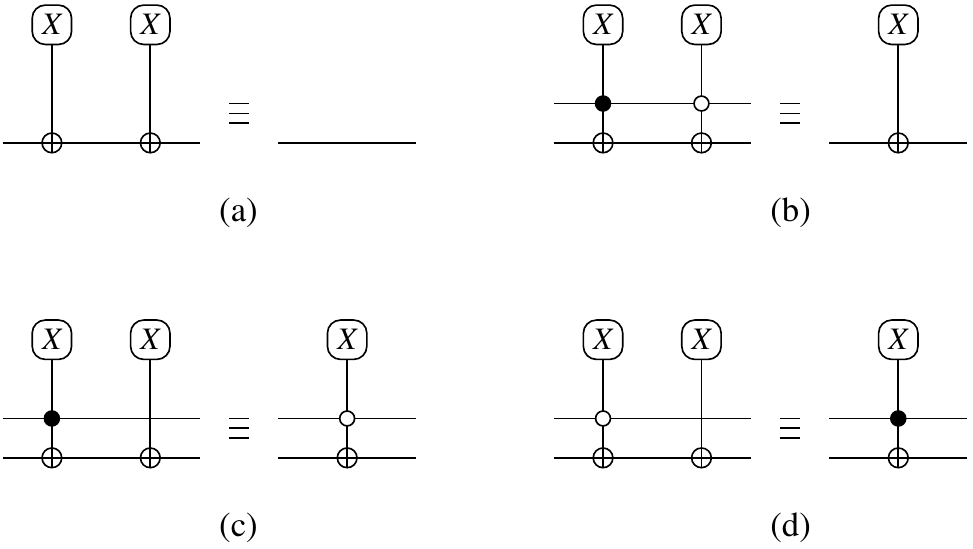}
  \caption{Simplification rules for gates acting on the same target
    bit. (a) Annihilation. (b) Control bit elimination. (c) and (d)
    Control bit reversal.}
  \label{fig:simplifications}
\end{figure}

In addition to the collisions in
Figs.~\ref{fig:collisions_combined}(c)
and~\ref{fig:collisions_combined}(d), which proliferate the number of
gates through the addition of debris, there are also simplifying
collisions resulting in complete or partial annihilation and thus a
reduction in the number of gates. For instance, as seen in
Fig.~\ref{fig:simplifications}, two gates can annihilate each other
completely, if they are identical, or combine into a single gate if
they only differ by one control bit.


\begin{figure}[h]
  \centering
  \includegraphics[angle=0,scale=0.75]{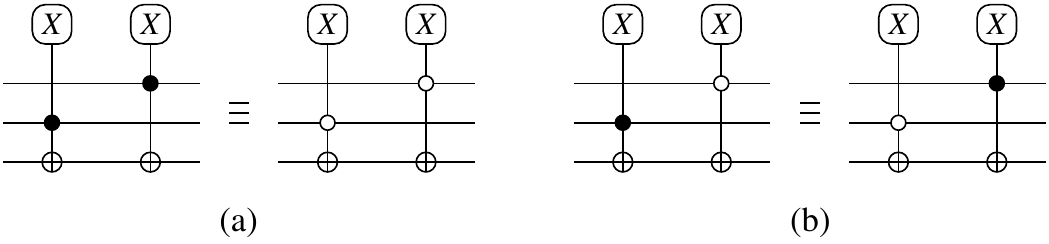}
  \caption{Additional rules that allow for changing the type (negated
    or non-negated -- or color) of the control bits.}
  \label{fig:color-rules}
\end{figure}

There are also some collisions of commuting gates where it is possible to
change the polarity of the controls, from negated to non-negated and
vice-versa, as illustrated in Fig.~\ref{fig:color-rules}.

\subsection{Factorization rules}

\begin{figure}[h]
  \centering
  \includegraphics[angle=0,scale=0.75]{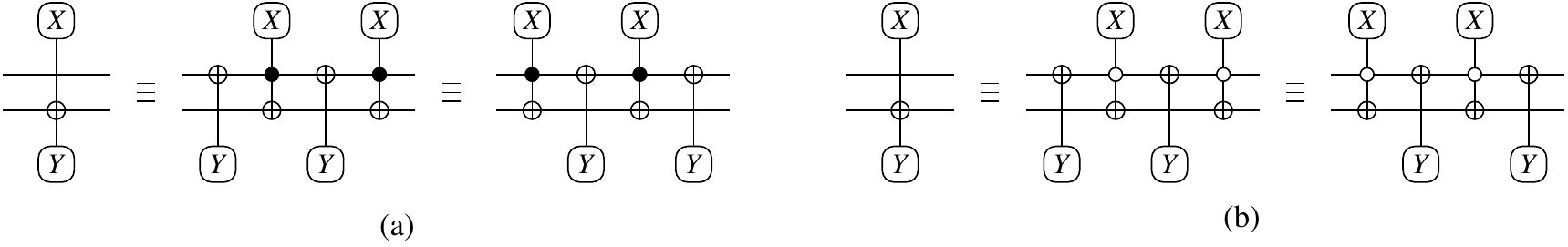}
  \caption{Factorization rules.}
    \label{fig:factorization-rules}
\end{figure}

Finally, we can factorize a gate with tail $XY$ into products of gates
with smaller tails by using the one-headed collision rules in
Fig.~\ref{fig:collisions_combined}(c), as illustrated in
Fig.~\ref{fig:factorization-rules}. Notice that the factorization of
the tail $XY$ into $X$ and $Y$ pieces may be done in several different
ways, and that there is also freedom in choosing the bit line on which
an additional control is placed.

\section{Rules for conjugation by inflationary gates}
\label{sec:conjugation_inflationary_gates}

\begin{figure}[h]
  \centering
  \vspace{0.5cm}
  \includegraphics[angle=0,scale=1.2]{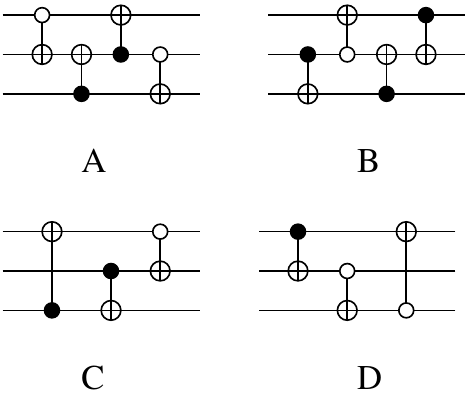}
  \vspace{0.5cm}
  \caption{Inflationary 3-bit gates expressed in terms of CNOTs (from
    Ref.~\cite{cipher-paper}). By permuting bitlines and control
    polarities, one obtains 24 distinct inflationary gates from
    topology A, 24 from B, 48 from C, and 48 from D, for a
    total of 144.}
  \label{fig:inflationary-gates}
\end{figure}

\begin{figure}[h]
  \centering
  \vspace{0.5cm}
  \includegraphics[angle=0,scale=0.80]{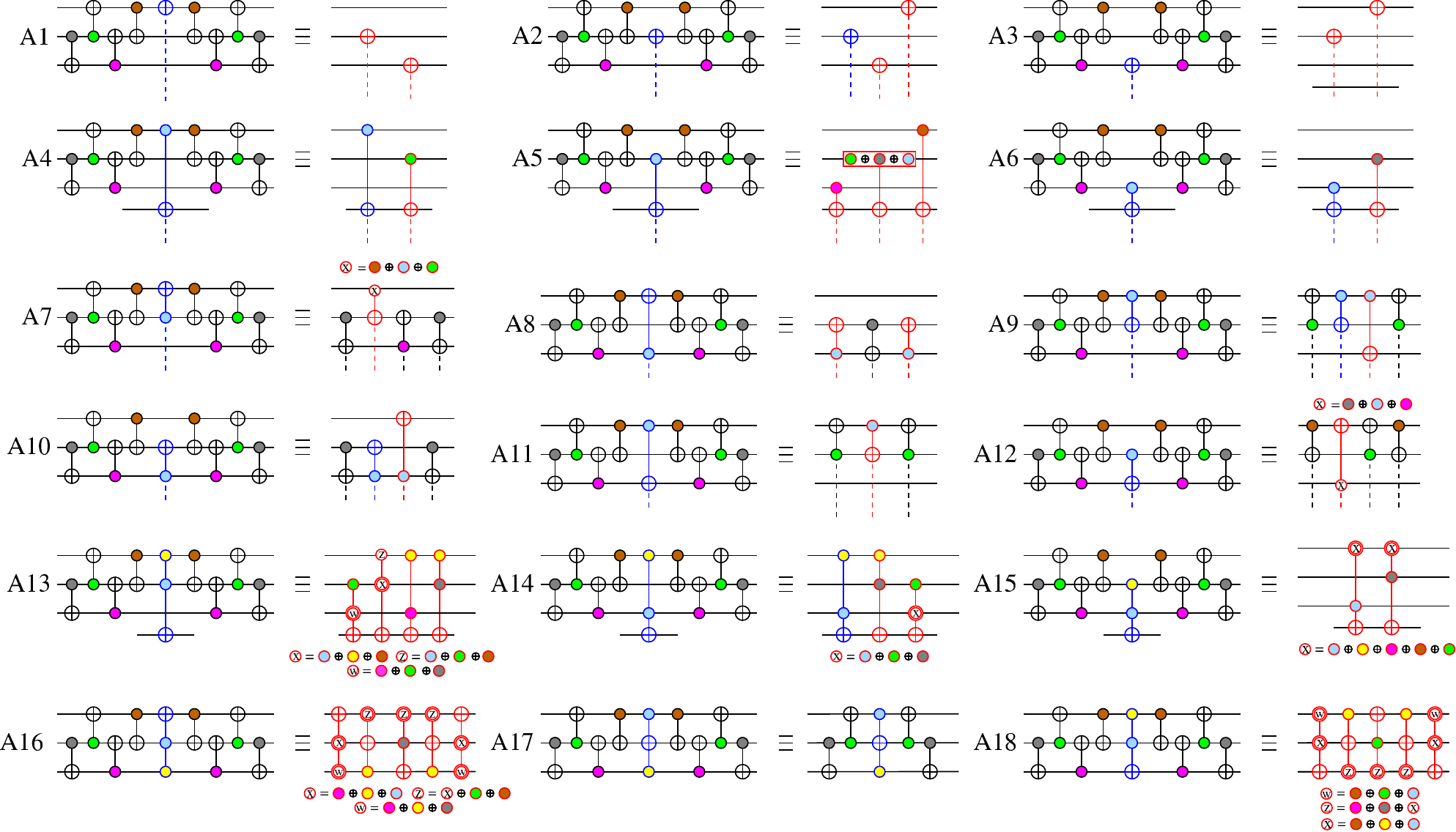}
  \vspace{0.5cm}
  \caption{Rules for conjugation with the inflationary gates in class
    A of Fig.~\ref{fig:inflationary-gates}. Each case comprises two
    circuits: pre- and post-conjugation (left and right,
    respectively). The gate being conjugated is shown with blue lines
    on the left circuit, surrounded by the inflationary gate block and
    its inverse (CNOT gates with black lines). The controls in the
    inflationary block gates have different colors so that their
    influence on the polarities of controls of offspring gates can be
    tracked down. The offspring gates on the right circuits have
    either blue lines (when they correspond to the original gate being
    conjugated) or red lines (when they are new gates). The dashed
    lines indicate connections that gates may have to additional
    bitlines. In a case A5 the polarity of the control of an offspring
    gate depends on the relative polarity of three controls (e.g., two
    from the inflationary gate and one from the gate being
    conjugated), with the minority polarity winning. In cases A7, A12,
    A13, A14, A15, A16, and A18, when a control polarity depends on
    polarities from multiple pre-conjugation gates, a polarity
    variable ``x'' (or ``w'' or ``z'') is inserted and defined below
    the circuit where it is utilized. Circuits resulting from
    conjugations are not necessarily unique and other equivalent
    circuits are possible. We choose those with the smallest number of
    gates and which minimize the the appearance of pre-conjugation
    gates.}
  \label{fig:conj_general1}
\end{figure}

\begin{figure}[h]
  \centering
  \vspace{0.5cm}
  \includegraphics[angle=0,scale=0.80]{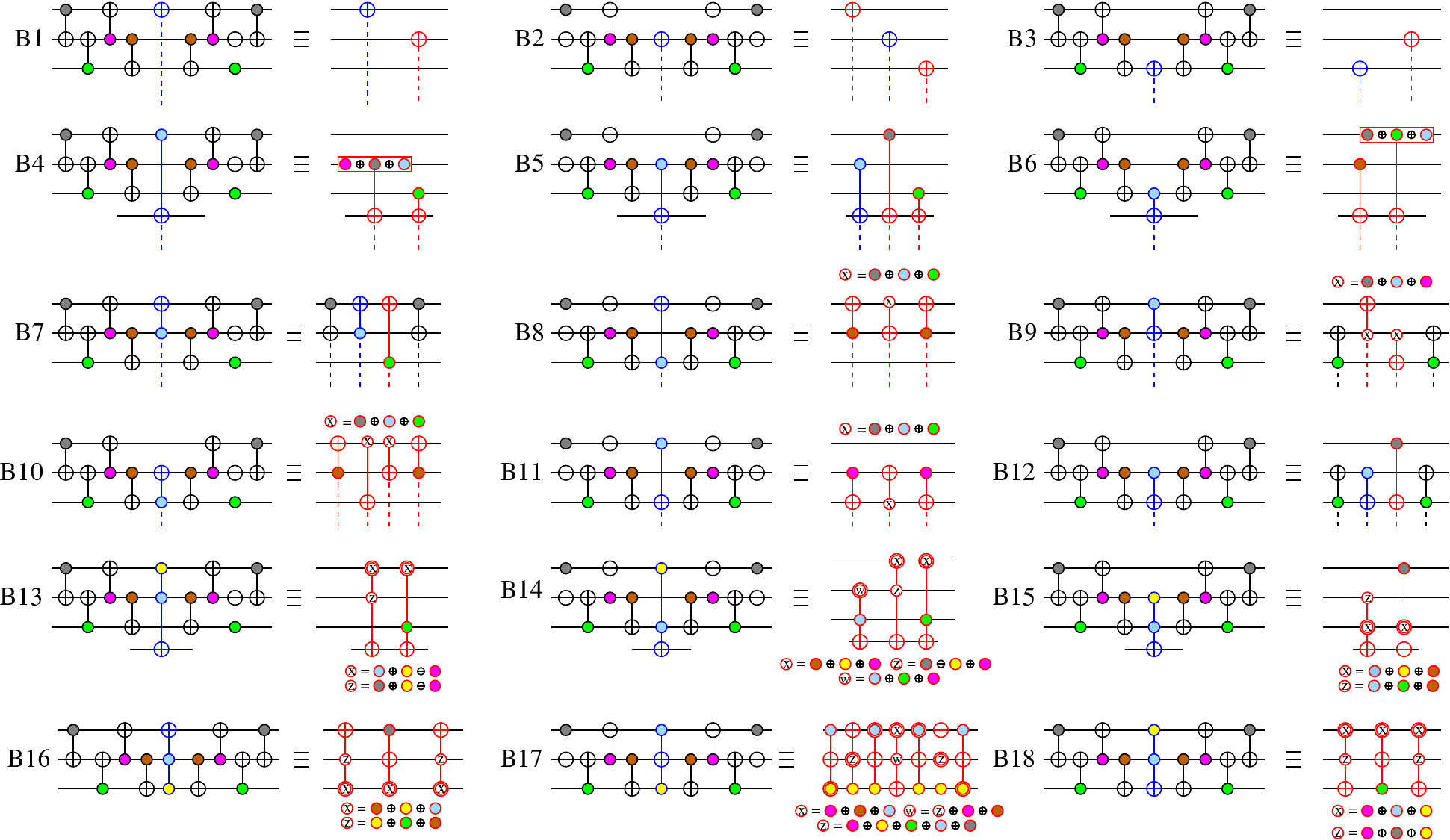}
  \vspace{0.5cm}
  \caption{Rules for conjugation with the inflationary gates in class
    B of Fig.~\ref{fig:inflationary-gates}. The same notation and
    conventions of Fig. \ref{fig:conj_general1} apply here.}
  \label{fig:conj_general2}
\end{figure}

\begin{figure}[h]
  \centering
  \vspace{0.5cm}
  \includegraphics[angle=0,scale=0.80]{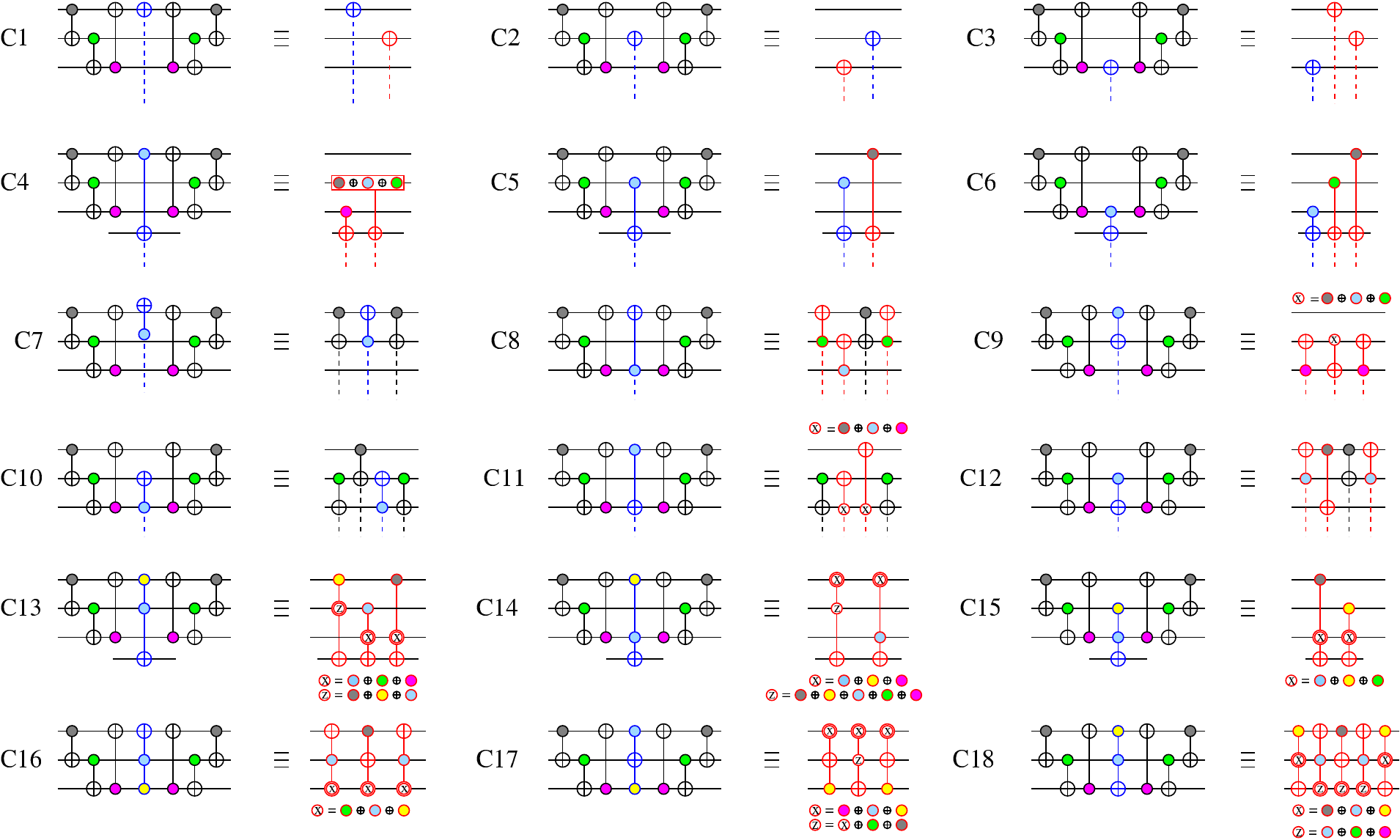}
  \vspace{0.5cm}
  \caption{Rules for conjugation with the inflationary gates in class
    C of Fig.~\ref{fig:inflationary-gates}. The same notation and
    conventions of Fig. \ref{fig:conj_general1} apply here.}
  \label{fig:conj_general3}
\end{figure}

\begin{figure}[h]
  \centering
  \vspace{0.5cm}
  \includegraphics[angle=0,scale=0.80]{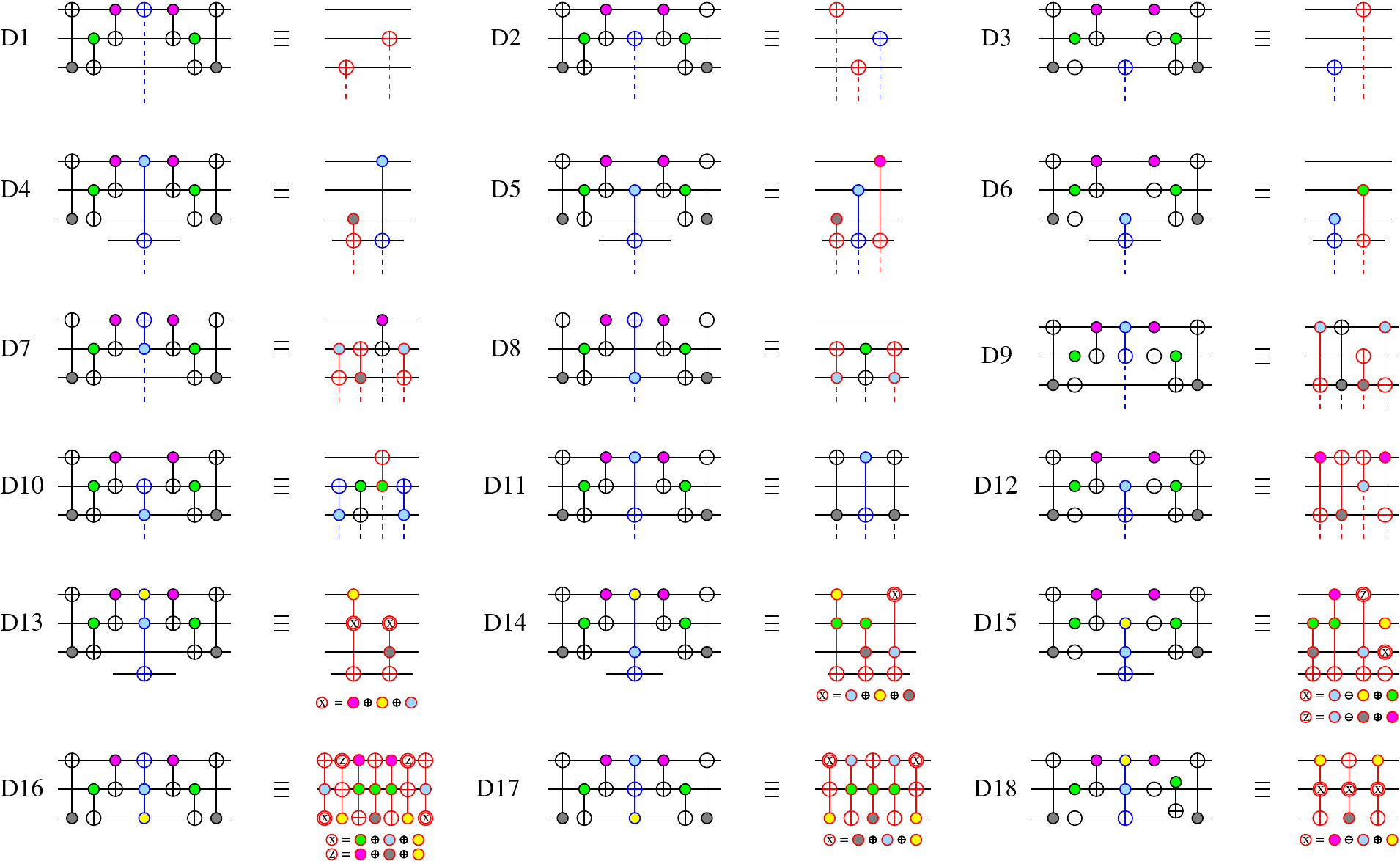}
  \vspace{0.5cm}
  \caption{Rules for conjugation with the inflationary gates in class
    D of Fig.~\ref{fig:inflationary-gates}. The same notation and
    conventions of Fig. \ref{fig:conj_general1} apply here.}
  \label{fig:conj_general4}
\end{figure}

In this section we present the rules for conjugation of controlled
gates by inflationary gates, which are depicted in
Fig.~\ref{fig:inflationary-gates}. There are four topologies, labeled
A, B, C and D, which are used in categorizing the conjugation rules
that are explained in pictures in Figs.~\ref{fig:conj_general1},
\ref{fig:conj_general2}, \ref{fig:conj_general3}, and
\ref{fig:conj_general4}.

\clearpage 

  
\clearpage 


\clearpage 


\clearpage

\section{Bound on BDD sizes using a linear network model}
\label{linear_network_model_bound}

The scaling of the size of the BDDs associated to the $3^\ell$ outputs
of the chip after $\ell$ layers of conjugation can also be obtained
recursively deploying the linear network model explained in
Refs.~\cite{MacMillan,Bryant_review,Knuth-book}. Here we rederive the bound
in Eq.~\eqref{eq:max_size_ell} using this method.

\begin{figure}[h]
  \centering
  \vspace{0.5cm}
  \includegraphics[angle=0,scale=0.26]{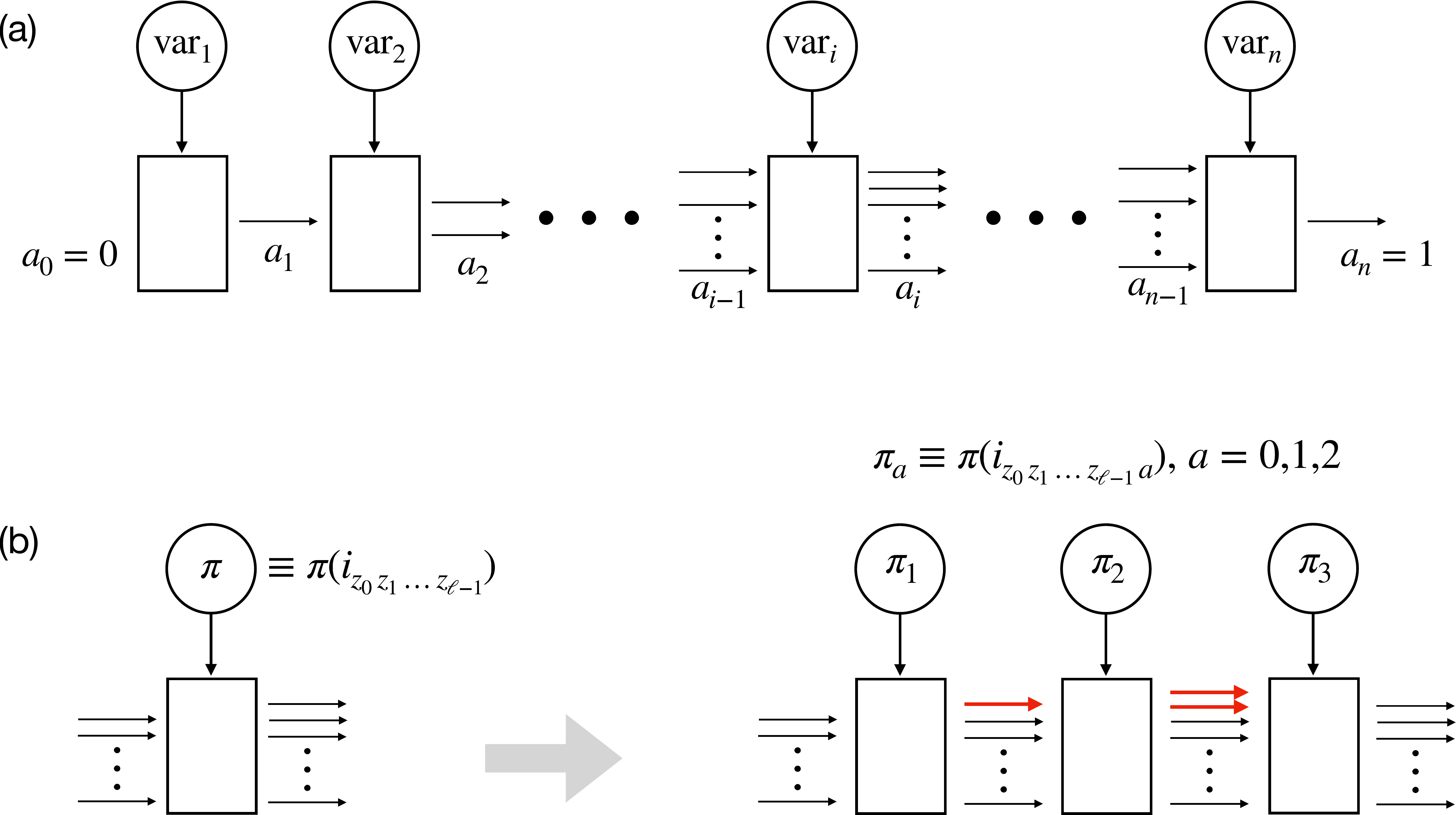}
  \vspace{0.5cm}
  \caption{(a) Linear network model of computation used in placing
    bounds on BDD
    sizes~\cite{MacMillan,Bryant_review,Knuth-book}. Each of of the
    modules (square boxes) receive bits of information from the module
    on its left (horizontal arrows) and one bit from the input
    variable above the module (vertical downward arrow), processes
    this information and passes signals to the next module on its
    right. (b) Linear network model of computation counterpart to the
    substitution depicted in Fig.~\ref{fig:BDD_recursion}. The module
    with input indexed by $\pi(i_{z_0z_1\dots z_{\ell-1}})$ is
    substituted by modules with inputs labeled by $\pi(i_{z_0z_1\dots
      z_{\ell-1}z_\ell})$, $z_\ell=0,1,2$, that represent a function
    $g^{-1}_{\pi(i_{z_0z_1\dots z_{\ell-1}})}$ of three variables
    $x_{\pi(i_{z_0z_1\dots z_{\ell-1}z_\ell})}$, $z_\ell=0,1,2$. The
    red arrows represent the additional bits of information that must
    be passed between the modules to carry out the computation of the
    function. The worst-case is depicted, in which one additional bit
    must be passed between the first and second modules, and two
    additional bits must be passed between the second and third
    modules.}
  \label{fig:linear_network_model}
\end{figure}

After the first layer of conjugation, the three outputs
$h_{\pi(i_0)},h_{\pi(i_1)}$ and $h_{\pi(i_2)}$ of the chip can be
encoded each in a BDD with three inputs, $x_{\pi(i_0)},x_{\pi(i_1)}$
and $x_{\pi(i_2)}$. We can represent the Boolean expression for each
of the outputs, for instance $h_{\pi(i_0)}$, as the result from a
linear network model of computation~\cite{Bryant_review,Knuth-book},
which is a useful representation that allows us to put bounds on the
sizes of the BDDs, as we show
below. Fig.~\ref{fig:linear_network_model}a depicts the linear
network model. Each of the modules (square boxes) takes a signal
through wires, each representing a bit of information, from the module
on its left and one signal from the input variables on top of the
module. Any function of 3 bits can be computed by such a linear
network model of computation, using 3 modules. In the worst case
scenario, the linear network operates as follows: (1) the information
on the first input variable is passed from the first to the second
module using a single wire; (2) the information on the first two
variables is passed from the second to the third module using two
wires; and, finally (3) the Boolean function of 3 variables is
computed by the third module with the information from the previous
two modules. (We note that this computation is carried solely by
moving information unidirectionally from one module to the next.)


The BDDs resulting from conjugation by the second layer can be
constructed following the prescription in Sec.~\ref{sec:nonlinear}. For
example, for the conjugation with the gates ($g$ and $g^{-1}$) that
overlap with bit ${\pi(i_0)}$, we first obtain the BDD for
$\tilde h_{\pi(i_0)}$ following the first step described in
Eq.~\eqref{eq:step1}. The tree structure of the fully-packed cipher
circuit implies that two fresh variables will be added, those that
join ${\pi(i_0)}$ in the triplet of bits acted on by $g$ and
$g^{-1}$. One can extend the linear network model by local
restructuring of the module for which $x_{\pi(i_0)}$ is the input, as
shown in Fig.~\ref{fig:linear_network_model}b, breaking it into 3
modules, each with input variables $x_{\pi(i_{0z_1})}$,
$z_1=0,1,2$. These three variables are those corresponding to the bits
in a triplet with ${\pi(i_0)}$ (one of them is $x_{\pi(i_0)}$
itself). This splitting of one module into three only requires the
addition of wires running internally between those three modules (see
Fig.~\ref{fig:linear_network_model}b): one wire running from the first
module to the second, and two wires from the second to the
third. These internal wires contain the additional information to
calculate the output of $g^{-1}$ needed for the substitution of
$x_{\pi(i_0)}\leftarrow g_{\pi(i_0)}^{-1}(x_{\pi(i_{00})},
x_{\pi(i_{01})}, x_{\pi(i_{02})})$, see Eq.~\eqref{eq:step1}. In a
similar fashion, we restructure the modules with input variables
$x_{\pi(i_1)}$ and $x_{\pi(i_2)}$. The result is a linear network
model to compute $\tilde h_{\pi(i_0)}$ given nine input variables,
$x_\pi(i_{z_0z_1}), z_0,z_1=0,1,2$.

We remark that while $\tilde h_{\pi(i_0)}$ depends on 9 variables, the
$\tilde h$ variable associated with the other two fresh bits that form
the triplet with ${\pi(i_0)}$ (two of ${\pi(i_{00})}, {\pi(i_{01})}$,
and ${\pi(i_{02})}$, with the third being ${\pi(i_0)}$ itself) simply
equals the corresponding output bits from $g^{-1}(x_{\pi(i_{00})},
x_{\pi(i_{01})}, x_{\pi(i_{02})})$.

The next step is to implement the calculation of
$h^g_{\pi(i_0)}(x)= g_{\pi(i_0)}\left(\tilde h_{\pi(i_0)},\tilde
  h_{\pi(i_1)},\tilde h_{\pi(i_2)}\right)$, as prescribed in
Eq.~\eqref{eq:step2}. If one arranges $x_{\pi(i_0)}$ to appear as
input to the rightmost module (this can always be done for each
separate output of the chip), no additional wires between the modules
need to be added in order to compute $h^g_{\pi(i_0)}$ from
$\tilde h_{\pi(i_0)}$.

These steps can be repeated recursively for conjugation with the
subsequent layers: a module from the previous layer is broken down
into three consecutive modules. In addition to the wires entering (on
the left) and exiting (on the right) the original module, one must add
wires between those three new modules: a wire from the first to the
second module, and two from the second to the third, as depicted in
Fig.~\ref{fig:linear_network_model}c. This recursion results in a
relation between the number of wires $a^{(\ell)}_m$ connecting
consecutive modules, labeled $m-1$ and $m$, of the linear network
obtained after conjugations with $\ell$ layers of nonlinear gates:
\begin{align}
  a^{(\ell+1)}_{3m} &= a^{(\ell)}_m\nonumber\\
  a^{(\ell+1)}_{3m+1} &= a^{(\ell)}_m+1\nonumber\\
  a^{(\ell+1)}_{3m+2} &= a^{(\ell)}_m+2
  \;.
  \label{eq:wires}
\end{align}

The size of the BDDs can be bounded given the number of wires
$a^{(\ell)}_m$ connecting consecutive modules, labeled $m-1$ and $m$,
of the linear network~\cite{Bryant_review,Knuth-book}:
\begin{align}
B\le B_{\rm max}(\ell)=\sum_{m=0}^{3^\ell} 2^{a^{(\ell)}_m}
  =\sum_{m=0}^{3^\ell-1} 2^{a^{(\ell)}_m}+2
  \;,
\end{align}
where we used that the last module has a single output wire,
$a^{(\ell)}_{3^\ell}=1$. Using the recursion for the number of wires
\begin{align}
  B_{\rm max}(\ell+1)-2
  &=
    \sum_{p=0}^{3^{\ell+1}-1}\; 2^{a^{(\ell+1)}_p}
    \nonumber\\
  &=
  \sum_{m=0}^{3^{\ell}-1}\; \left(2^{a^{(\ell+1)}_{3m}}+2^{a^{(\ell+1)}_{3m+1}}+2^{a^{(\ell+1)}_{3m+2}} \right)
  \nonumber\\
  &=
  \sum_{m=0}^{3^{\ell}-1}\; \left(2^{a^{(\ell)}_m}+2^{a^{(\ell)}_m+1}+2^{a^{(\ell)}_m+2} \right)
  \nonumber\\
  &=
  \sum_{m=0}^{3^{\ell}-1}\; 2^{a^{(\ell)}_m}\;\left(1+2+2^2\right)
  \nonumber\\
  &=
  7\;\left[B_{\rm max}(\ell)-2\right]
  \;.
  \label{eq:max_size_recursion_appendix}
\end{align}
Seeding the recursion with $B_{\rm max}(0)=3$ (the size of the BDD
representing a simple NOT operation) yields
\begin{align}
  B_{\rm max}(\ell)
  =
  7^\ell + 2
  \;,
  \label{eq:max_size_ell_appendix}
\end{align}
the same result as in Eq.~\eqref{eq:max_size_ell}.


\bibliographystyle{splncs04}
\bibliography{references}

\begin{thebibliography}{10}
\providecommand{\url}[1]{\texttt{#1}}
\providecommand{\urlprefix}{URL }
\providecommand{\doi}[1]{https://doi.org/#1}

\bibitem{Barak}
Barak, B., Goldreich, O., Impagliazzo, R., Rudich, S., Sahai, A., Vadhan, S.,
  Yang, K.: On the (im)possibility of obfuscating programs. In: Kilian, J.
  (ed.) Advances in Cryptology --- CRYPTO 2001. pp. 1--18. Springer Berlin
  Heidelberg, Berlin, Heidelberg (2001)

\bibitem{Bryant1986}
Bryant, R.: Graph-based algorithms for {Boolean} function manipulation. IEEE
  Transactions on Computers  \textbf{35}(08),  677--691 (aug 1986).
  \doi{10.1109/TC.1986.1676819}

\bibitem{Bryant_review}
Bryant, R.E.: Symbolic {Boolean} manipulation with ordered binary-decision
  diagrams. ACM Computing Surveys  \textbf{24}(3),  293--318 (Sep 1992).
  \doi{10.1145/136035.136043}, \url{https://doi.org/10.1145/136035.136043}

\bibitem{cipher-paper}
Chamon, C., Mucciolo, E.R., Ruckenstein, A.E.: Quantum statistical mechanics of
  encryption: Reaching the speed limit of classical block ciphers. Annals of
  Physics  \textbf{446},  169086 (2022).
  \doi{https://doi.org/10.1016/j.aop.2022.169086}

\bibitem{TFHE}
Chillotti, I., Gamma, N., Georgieva, M., Izabachene, M.: {TFHE}: Fast fully
  homomorphic encryptionover the torus. J. Cryptol.  \textbf{33},  34--91
  (2020)

\bibitem{Chillotti2020}
Chillotti, I., Gama, N., Georgieva, M., Izabach{\`e}ne, M.: {TFHE}: Fast fully
  homomorphic encryption over the torus. J. Cryptology  \textbf{33}(1),  34--91
  (Jan 2020)

\bibitem{Fredkin1982}
Fredkin, E., Toffoli, T.: Conservative logic. International Journal of
  Theoretical Physics  \textbf{21}(3-4),  219--253 (Apr 1982).
  \doi{10.1007/bf01857727}, \url{https://doi.org/10.1007/bf01857727}

\bibitem{FV-NFLib}
{FV-NFLib}. \url{https://github.com/CryptoExperts/FV-NFLlib} (Jul 2016)

\bibitem{gentry2009}
Gentry, C.: Fully homomorphic encryption using ideal lattices. In: Proceedings
  of the Forty-First Annual ACM Symposium on Theory of Computing. p. 169–178.
  STOC '09, Association for Computing Machinery, New York, NY, USA (2009).
  \doi{10.1145/1536414.1536440}, \url{https://doi.org/10.1145/1536414.1536440}

\bibitem{Goldwasswer-Rothblum}
Goldwasser, S., Rothblum, G.N.: On best-possible obfuscation. In: Vadhan, S.P.
  (ed.) Theory of Cryptography. pp. 194--213. Springer Berlin Heidelberg,
  Berlin, Heidelberg (2007)

\bibitem{Harrow2018approximate}
Harrow, A., Mehraban, S.: Approximate unitary $t$-designs by short random
  quantum circuits using nearest-neighbor and long-range gates (2018)

\bibitem{HEAAN}
{HEAAN} {v.2.1}. \url{https://github.com/snucrypto/HEAAN} (Sep 2018)

\bibitem{IBM-HELib}
{IBM} {HElib} (v2.2.1). \url{https://github.com/homenc/HElib} (Oct 2021), {IBM}
  Research, Europe

\bibitem{Iwama2002}
Iwama, K., Yamashita, S.a.: Transformation rules for {CNOT}-based quantum
  circuits and their applications. New Gener. Commput.  \textbf{21},  297--317
  (2003). \doi{10.1007/BF03037305}

\bibitem{Knuth-book}
Knuth, D.E.: The Art of Computer Programming, vol.~4. Addison-Wesley (2019)

\bibitem{cryptoeprint_32_64_bit}
Lee, S., Shin, D.J.: Overflow-detectable floating-point fully homomorphic
  encryption. Cryptology ePrint Archive, Paper 2022/186 (2022),
  \url{https://eprint.iacr.org/2022/186},
  \url{https://eprint.iacr.org/2022/186}

\bibitem{RLWE}
Lyubashevsky, V., Peikert, C., Regev, O.: On ideal lattices and learning with
  errors over rings. J. ACM  \textbf{60}(6) (nov 2013). \doi{10.1145/2535925},
  \url{https://doi.org/10.1145/2535925}

\bibitem{naor-reingold}
Naor, M., Reingold, O.: Synthesizers and their application to the parallel
  construction of pseudo-random functions. Journal of Computer and System
  Sciences  \textbf{58}(2),  336--375 (1999).
  \doi{https://doi.org/10.1006/jcss.1998.1618}

\bibitem{nielsen2002quantum}
Nielsen, M.A., Chuang, I.: Quantum computation and quantum information.
  Cambridge University Press, Cambridge, UK (2010)

\bibitem{Palisade}
{Palisade}. \url{https://palisade-crypto.org/software-library} (May 2021)

\bibitem{MacMillan}
Ravi, K., McMillan, K.L., Shiple, T.R., Somenzi, F.: Approximation and
  decomposition of binary decision diagrams. In: Proceedings of the 35th Annual
  Design Automation Conference. p. 445–450. DAC '98, Association for
  Computing Machinery, New York, NY, USA (1998). \doi{10.1145/277044.277168},
  \url{https://doi.org/10.1145/277044.277168}

\bibitem{Regev}
Regev, O.: On lattices, learning with errors, random linear codes, and
  cryptography. In: Proceedings of the Thirty-Seventh Annual ACM Symposium on
  Theory of Computing. p. 84–93. STOC '05, Association for Computing
  Machinery, New York, NY, USA (2005). \doi{10.1145/1060590.1060603},
  \url{https://doi.org/10.1145/1060590.1060603}

\bibitem{Regev-and-Co}
Regev, O.: On lattices, learning with errors, random linear codes, and
  cryptography. J. ACM  \textbf{56}(6) (Sep 2009).
  \doi{10.1145/1568318.1568324}, \url{https://doi.org/10.1145/1568318.1568324}

\bibitem{Roberts2015}
Roberts, D.A., Stanford, D., Susskind, L.: Localized shocks. JHEP
  \textbf{2015}(3) (Mar 2015). \doi{10.1007/jhep03(2015)051},
  \url{https://doi.org/10.1007/jhep03(2015)051}

\bibitem{MS-Seal}
{M}icrosoft {SEAL} (release 3.6). \url{https://github.com/Microsoft/SEAL} (Nov
  2020), microsoft Research, Redmond, WA.

\bibitem{Smart2014}
Smart, N.P., Vercauteren, F.: Fully homomorphic {SIMD} operations. Des. Codes
  Cryptogr.  \textbf{71}(1),  57--81 (Apr 2014)

\end{thebibliography}


\end{document}